\documentclass[preprint,authoryear,12pt]{elsarticle_preprint}
\usepackage[svgnames]{xcolor}
\usepackage{graphicx, multirow, subfigure}
\usepackage{amsmath, amssymb}
\usepackage{pdflscape}
\journal{Solar Energy}
\begin{document}

\begin{frontmatter}
\title{A semiparametric spatio-temporal model \\ for solar irradiance data}
\author{Joshua D.~Patrick\corref{cor1}}
\address{Department of Statistics,
One Shields Avenue, University of California, 
Davis, CA 95616-5270}
\ead{jdpatrick@ucdavis.edu}
\cortext[cor1]{Corresponding author}
\author{Jane L.~Harvill}
\address{Department of Statistical Science,
P.O.~Box 97140,
Baylor University,
Waco, TX 76798-7140}
\ead{jane\_harvill@baylor.edu}
\author{Clifford W.~Hansen}
\address{P.O.~Box 5800, 
Sandia National Laboratories, 
Albuquerque, NM 87185-1033}
\ead{cwhanse@sandia.gov}

\begin{abstract}
Design and operation of a utility scale photovoltaic (PV) power plant
depends on accurate modeling of the power generated, which is highly
correlated with aggregate solar irradiance on the plant's PV
modules.  At present, aggregate solar irradiance over the area of a typical PV power plant cannot be measured directly.  Rather, irradiance
measurements are typically available from a few, relatively small sensors and thus
aggregate solar irradiance must be estimated from these data.  As a step towards finding more accurate methods for estimating aggregate irradiance from avaialble measurements, we evaluate semiparametric spatio-temporal models for global
horizontal irradiance.  Using data from a 1.2 MW PV plant located in Lanai, Hawaii, we show that a semiparametric model can be more accurate than simple intepolation between sensor locations.  We investigate spatio-temporal models with separable and nonseparable covariance structures and find no evidence to support assuming a separable covariance structure.
\end{abstract}

\begin{keyword}
Irradiance \sep Models \sep Spatio-temporal model \sep Nonseparability \sep
Lattice data \sep Semiparametric time series
\end{keyword}
\end{frontmatter}
\section{Introduction}
Accurate modeling of power output from utility scale photovolaic (PV)
power plants is often key to obtaining favorable financial terms
during system design and construction, and to efficient and profitable
plant operation.  Accurate modeling of power output requires estimating 
aggregate plane-of-array (POA)  irradiance over the plant's footprint with sufficient
precision, because aggregate POA irradiance is highly
correlated with power output \citep{kuszamaul2010}.  

There is great interest in methods to improve the accuracy of estimates of aggregate 
irradiance used for PV power plant modeling.  Error in estimating the aggregate
irradiance translates directly to error in modeled power output and
thus to error in projected energy production; relatively small errors
in projected energy may translate to significant uncertainty in
projected profit because utility-scale
PV plants are typically leveraged financially.

Here, we explore statistical modeling of global horizontal irradiance
(GHI) at spatial and temporal scales relevant to design and operation
of a utility-scale PV power plant, i.e., on the order of 1
${\mbox{km}}^2$ and a few minutes.  We apply recent advances in
spatio-temporal statistical methods and illustrate our results with
data from a 1.2MW PV plant at La Ola, Lanai, HI.  We pursue
semiparametric (i.e., data-driven) rather than parametric approaches
because a successful model could then be applied regardless of weather
conditions at the location of interest.  In contrast, parametric
models implicitly assume that random variables in the model (e.g.,
GHI) are well-described by specified distributions, an assumption
which may not hold if weather conditions change.  We compare the 
resulting models with the commonly used simple spatial average 
which estimates aggregate irradiance over a plant's footprint by 
averaging measurements from sensors located in or near the plant. 

A challenge in modeling irradiance data is incorporating the
interaction between time and space.  Especially in the presence of
advecting clouds, the irradiance observed at one location is likely to
also be observed at other locations but with a time shift.  Thus we
anticipate that irradiance will exhibit a spatial autocorrelation that
varies with time.  Spatio-temporal models explicitly account for 
this autocorrelation and thus may predict aggregate irradiance 
more accurately than does a simple spatial average.

Fundamentally, GHI can be viewed as a random spatio-temporal process.
Literature reports several efforts at modeling individual time-series
of irradiance, and substantially fewer attempts to construct
spatio-temporal models of irradiance considering several proximal
locations.  The literature on individual time-series modeling includes approaches
based on autoregressive integrated moving average (ARIMA) analysis
\citep[e.g.,][]{yang2012}, non-linear autogressive analysis \citep
      {glasbey2001}, regression analysis \citep{Reikard2009},
      artificial neural networks analysis \citep{paoli2010},
      $k$-Nearest Neighbors algorithm \citep{paoli2010} and Bayesian
      inference \citep{paoli2010}.  These approaches focus on
      forecasting irradiance considering only the measurements at a
%
%
      selected location separately from measurements at other
      locations.  \cite{paoli2010} considers a type of artificial
      neural network known as Multi-Layer Perceptron (MLP) network and
      finds their method performs as well or better than other methods
      such as ARIMA analysis, Bayesian inference, and $k$-Nearest
      Neighbors.  \cite{yang2012} introduce an ARIMA model that
      incorporates low-resolution, ground-based cloud cover data to
      obtain next hour solar irradiance. The authors state that their
      ARIMA model outperforms all other time series forecasting
      methods in four of the six stations they tested.  In both the
      MLP and ARIMA methods, the model does not incorporate a spatial
      component but only models irradiance in time.

The literature on spatio-temporal modeling of solar irradiance is
limited. To our knowledge, ours is the first attempt to model
irradiance at time and spatial scales relevant to modeling a
utility-scale PV plant.  \cite{glasbey2008} model irradiance data
from ten sensors roughly at 5km spacing using a spatio-temporal
autoregressive moving average (STARMA) model. The STARMA model
incorporates the Euclidian distances between two points in order to
model the spatial structure of the data. However, the STARMA model
used in \cite{glasbey2008} assumes a separable covariance structure,
an assumption which we find to be questionable at the scale of a
single PV plant.

We propose a spatio-temporal model that incorporates a data-driven
method for modeling the time series component.  Our model improves
upon the works of \cite{yang2012} and \cite{glasbey2001} because we do
not assume a parametric form for the time component of the model, and
improves on \cite{glasbey2008} through the nonseparable covariance
structure.  The remainder of this paper is organized as follows. In
Section~\ref{sec:timeseries}, we discuss how the time series structure
is modeled via a semiparametric model fitted with a data-driven method
known as spline-backfitted kernel (SBK) estimation.  In
Section~\ref{sec:spacetime}, we introduce the spatio-temporal model,
and compare the model's performance assuming either a separable or
a nonseparable covariance structure to evaluate whether separability
can be assumed.  In Section~\ref{sec:application}, we apply the model
to irradiance data from the La Ola photovoltaic plant in Lanai, HI.
Finally, we provide discussion and conclusions in
Section~\ref{sec:conclusion}.

\section{Modeling Irradiance}
\label{sec:timeseries}

 Let $Q_{s,t}$ represent an observable process,
e.g., measured GHI, at time $t$ and location $s$ for $t = 1, 2,
\ldots, T$ and $s = 1, 2, \ldots, S$.  If there is no interaction in
time and space, the covariance function of $Q_{s,t}$ can be written as
a product of two functions where one function is dependent on time
only, and the other on location alone.  Such a covariance function is
called ``separable.''  However, when interactions in space and time
are present, the covariance function is ``nonseparable;'' i.e., it
cannot be factored into two separate functions.  Spatio-temporal
models with separable covariance are much easier to implement.  But in
the presence of space-time interaction, separable models do not
perform well, and can lead to misleading or incorrect conclusions.

For modeling $Q_{s,t}$, consider
\begin{equation}
\label{eq:st1}
Q_{s,t} = R_{s,t} + Z_{s,t}, \quad t = 1, \ldots, T, \quad s = 1, \ldots, S,
\end{equation}
where, at time $t$ and location $s,~R_{s,t}$ represents the true
irradiance signal and $Z_{s,t}$ is a noise process.  Furthermore, decompose
the noise process into a sum of three terms,
\begin{equation}
\label{eq:st2}
Z_{s,t} = X_{s,t} + Y_{s,t} + \varepsilon_{s,t},
\end{equation}
where $X_{s,t}$ is a time series process at location $s$,
$Y_{s,t}$ is a spatial process at time $t$, and
$\varepsilon_{s,t}$ is a multivariate error process with mean zero and
$TS \times TS$ covariance matrix $\boldsymbol{\Sigma}(s,t)$.  If the
process is separable, then the covariance matrix can be written as
$\boldsymbol{\Sigma}(s,t) = \boldsymbol{\Lambda}(t) \otimes
\boldsymbol{\Gamma}(s)$, where $\boldsymbol{\Lambda}(t)$ is a $T
\times T$ temporal covariance matrix, $\boldsymbol{\Gamma}(s)$ is an
$S \times S$ spatial covariance matrix, and $\otimes$ is the Kronecker
product \citep{woolrich2004}.

There are a variety of methods for fitting separable spatio-temporal
models to space-time data.  A review of space-time analysis methods
and their computational counterparts can be found in
\cite{harvill2010} or \cite{cressiewikle2011}.  We consider three
approaches to fitting model~\eqref{eq:st2}.  The first approach fits a
spatial model at each time.  Then spatial residuals are computed, and
for each location a time series model is fitted to the spatial residuals at each location.  The second approach models the time series at each location, computes time residuals, and then fits a spatial model at each time to the residuals.  Both of these approaches carry the assumption that the covariance structure is separable in space and time.  The third approach removes the separability assumption, jointly modeling time and space using the spatio-temporal model introduced in Section~\ref{sec:FCSAR}.
 
\subsection{Modeling the Time Series Component}
\label{sec:FCAR}
For modeling time series data arising from a dynamic process, such as
solar irradiance, nonlinear models often out-perform linear models
\citep{tong1993}.  Although the class of nonlinear time series models
is infinitely large, there are many popular parametric nonlinear
models including the bilinear model \citep{sandg1984}, the exponential
autoregressive model \citep{hagganozaki1981}, and a variety of
threshold autoregressive models \citep{tong1983,vandijk1999}.  When
one of these parametric models is known to be appropriate for
analyzing the time series, it should be used for analyzing the
series.  However in the analysis of solar irradiance, no
specific class of parametric nonlinear model has been shown to be
generally applicable, and therefore we pursue a semiparametric
approach.  In this section, we examine only the time component of the
model.  So to ease notation, for the remainder of this section, we
consider the location fixed, and suppress the $s$ subscript; that is
$X_{s,t} = X_t$ for a fixed value of $s$.

A highly versatile semiparametric model is the functional coefficient
autoregressive model of order $p$ (FCAR($p$)), first introduced by
\cite{chen1993}.  The FCAR($p$) model has an additive autoregressive
structure, but with coefficients that vary as a function of some
variable, $u$ say, which can be exogeneous to the series $X_t$.  In
the pure time series context, $u$ is a lagged value of the series, and
we write $u_t = X_{t-d}$.  In this paper, we restrict the FCAR($p$)
models to those with $u_t = X_{t-d}$, and so define the FCAR($p$)
model as
\begin{equation}
\label{eq:fcar}
X_t = m_0(u_t) + \sum_{j=1}^p m_j(u_t)X_{t-j} + \omega_t, \qquad
t = p+1, \ldots, T
\end{equation}
where $u_t = X_{t-d},~d \le p,~m_j(\cdot),~j = 0, 1, 2, \ldots, p$
are measurable functions of $u$, and $\left\{\omega_t\right\}$ is a
sequence of independent and identically distributed (IID)~random
variables with mean zero and constant variance.

Reasonble use of the FCAR($p$) model requires only that the model is
additive, and places few restrictions on the functional coefficients.
To illustrate the versality of the FCAR($p$) model, note that if
$m_0(u_t) = 0$, and $m_j(u_t) = \alpha_j,~j = 1, 2, \ldots, p$ are
constants, then the FCAR($p$) reduces to a linear autoregressive model
of order $p,~X_t = \alpha_1 X_{t-1} + \cdots + \alpha_p X_{t-p} +
\omega_t$.  Another example is, for each $j = 1, \ldots, p$, the
coefficients are of the form $m_j(X_{t-d}) = \alpha_j + \beta_j
\exp\left\{-\delta X_{t-d}^2\right\}$.  Then the FCAR($p$) model
reduces to the exponential autoregressive model of
\cite{hagganozaki1981}.  Moreover, the FCAR($p$) formulation allows
for a mixture of models; for example, $m_1(X_{t-d}) = \alpha_1$ and
$m_2(X_{t-d}) = \alpha_2 + \beta_2\exp\left\{-\delta
X_{t-d}^2\right\}$.  \cite{fandy2003} contains a review of methods for
fitting the FCAR($p$) model, and related inferential procedures.  In
the following section, we propose a more recent, improved method for
fitting the FCAR($p$) model.

\subsection{Spline-Backfitting Kernel Estimation}
\label{sec:SBK}
With no presupposed form for the functional coefficients, we propose a
data-driven method for finding pointwise estimates of the functions
$m_j(u)$, $j = 0, 1, 2, \ldots, p$.  A number of methods are proposed
in the statistics literature.  \cite{chenliu2001} and
\cite{caietal2000} propose a kernel regression approach to fitting the
model. \cite{harvillray2006} extend the procedure to the case when the
series is a vector process.  More recently, spline-backfitted kernel
(SBK) estimation has been proposed as a means for fitting
semiparametric models like the FCAR($p$) model.  SBK estimation is an
adaptation of the backfitting algorithm of \cite{hastie1990}, and
combines the computational speed of splines with the asymptotic
properties of kernel smoothing.

The SBK method uses an under-smoothed centered standard spline
procedure to pre-estimate the $m_j(u),~j = 0, 1, 2, \ldots, p$.  These
pre-estimates, also called ``oracle'' estimates, are used to find
psuedo-responses.  Then the pseudo-responses are used to estimate the
$m_j(u)$ through a kernel estimator; e.g., the Nadaraya-Watson
estimator.  The SBK method was first proposed by \cite{wangyang2007}
for estimating nonlinear additive autoregressive models.
\cite{wangyang2009} adapt the SBK method for IID data, \cite{liu2011}
adapt it to generalized additive models, and \cite{ma2011} to
partially linear additive models.  \cite{liu2010} propose the SBK
method for additive coefficient models.

The ability to estimate $m_j(u),~j = 0, 1, 2, \ldots, p$ relies on the good
approximation properties of spline estimators.  For any $j = 0, 1, 2,
\ldots, p$, assume $m_j(\cdot)$ is sufficiently smooth.  Without
loss of generality, $u$ can be defined on the compact interval
$[0,1]$.  Define the integer $N \approx T^{2/5}\log T$, and let $H =
(N+1)^{-1}$.  Let $0 = \xi_0 < \xi_1 < \cdots < \xi_N < \xi_{N+1} = 1$
denote a sequence of equally spaced knots.
There is a set of basis functions $b_0(u), b_1(u), \ldots b_{N+1}(u)$
and a set of constants $\tilde\lambda_{0,j}, \tilde\lambda_{1,j},
\ldots, \tilde\lambda_{N+1,j}$ such that the spline estimator of the
$j$-th coefficient is
\begin{equation}
\label{eq:spline}
m_j(u) \approx \tilde m_j(u) = \sum_{k=0}^{N+1} \tilde\lambda_{k,j}b_k(u).
\end{equation}
For the basis functions, we choose the linear $B$-spline basis,
defined by 
$$
b_k(u) = \left(1 - \frac{|u - \xi_k|}{H}\right)_+ 
       = \left\{\begin{array}{ll}
               (N + 1)u - k + 1, & \xi_{k-1} \le u < \xi_k, \\
               k + 1 - (N + 1)u, & \xi_k < u \le \xi_{k+1}, \\
               0,                & {\mbox{otherwise.}}
               \end{array}\right.
$$ 
The coefficients $\tilde\lambda_{0,j}, \tilde\lambda_{1,j}, \ldots,
\tilde\lambda_{N+1,j}$ are estimated via least squares; that is, the
$\tilde\lambda_{k,j},~k = 0, 1, \ldots, N+1,~j = 0, 1, 2, \ldots, p$
are the values of $\lambda_{k,j}$ that minimize the sum of squares
\begin{equation}
\label{eq:SS}
\sum_{t=p+1}^T \left[X_t - 
 \sum_{j=1}^p \left\{\sum_{k=0}^{N+1}\lambda_{k,j} b_k(u)\right\}
  X_{t-j}\right]^2
\end{equation}

The spline-estimated functional coefficients are then used to compute
``pseudo-responses.''  Specifically, for each $j^\prime = 0, 1, 2, \ldots,
p, j^\prime \ne j$, the pseudo-responses are defined by
$$
\hat{W}_{t, j^\prime} = X_t - \sum_{j=1, j \ne j^\prime}^p \tilde m_j(u)X_{t-j}, 
    \quad t = p + 1, p + 2, \ldots, T.
$$ 
For each $j^\prime = 0, 1, 2, \ldots, p$, let
$\tilde{\textbf{W}}_{j^\prime} = (\tilde{W}_{p+1,{j^\prime}}, \cdots,
\tilde{W}_{T,{j^\prime}})^\prime$ represent the vector of
pseudo-responses, and define the matrix
$$
\textbf{M} = \text{diag}\left\{K_h \left(X_{p+1-d} - u\right),
  \ldots, K_h \left(X_{T-d} - u\right)\right\},
$$
where $K_h(\cdot) = h^{-1}K(\cdot/h),~K(\cdot)$ is a kernel function,
and $h > 0$ is a bandwidth.  Then the SBK estimator of $m_{j^\prime}(u)$ is
\begin{equation}
\label{eq:kernelfit}
\hat m_{j^\prime}\left(u\right) = 
  \left(\begin{array}{c}1 \\ 0\end{array}\right)\left(\frac{1}{T}
  \textbf{C}^\prime \textbf{M}\textbf{C}\right)^{-1}
  \frac{1}{T}\textbf{C}^\prime\textbf{M}\tilde{\textbf{W}}_{j^\prime},
\end{equation}
where 
$$
\textbf{C}^\prime = \left[\begin{array}{cccc}
  X_{p+1} & X_{p+2} & \cdots & X_T \\
  X_{p+1}\left(X_{p+1-d} - u\right) & X_{p+2}(X_{p+2-d} - u) & 
    \cdots & X_T\left(X_{T-d} - u\right)
  \end{array}\right]
$$
The idea behind SBK estimation is to under-smooth in the pre-estimates
in order to reduce the bias. This under-smoothing leads to a larger
variance which is reduced in the kernel estimation step. The use of
splines for the pre-estimates is computationally fast while using
kernel smoothing provides convenient asymptotic results \citep{liu2010}.

\begin{figure}
\begin{center}
\includegraphics[width=0.9\textwidth]{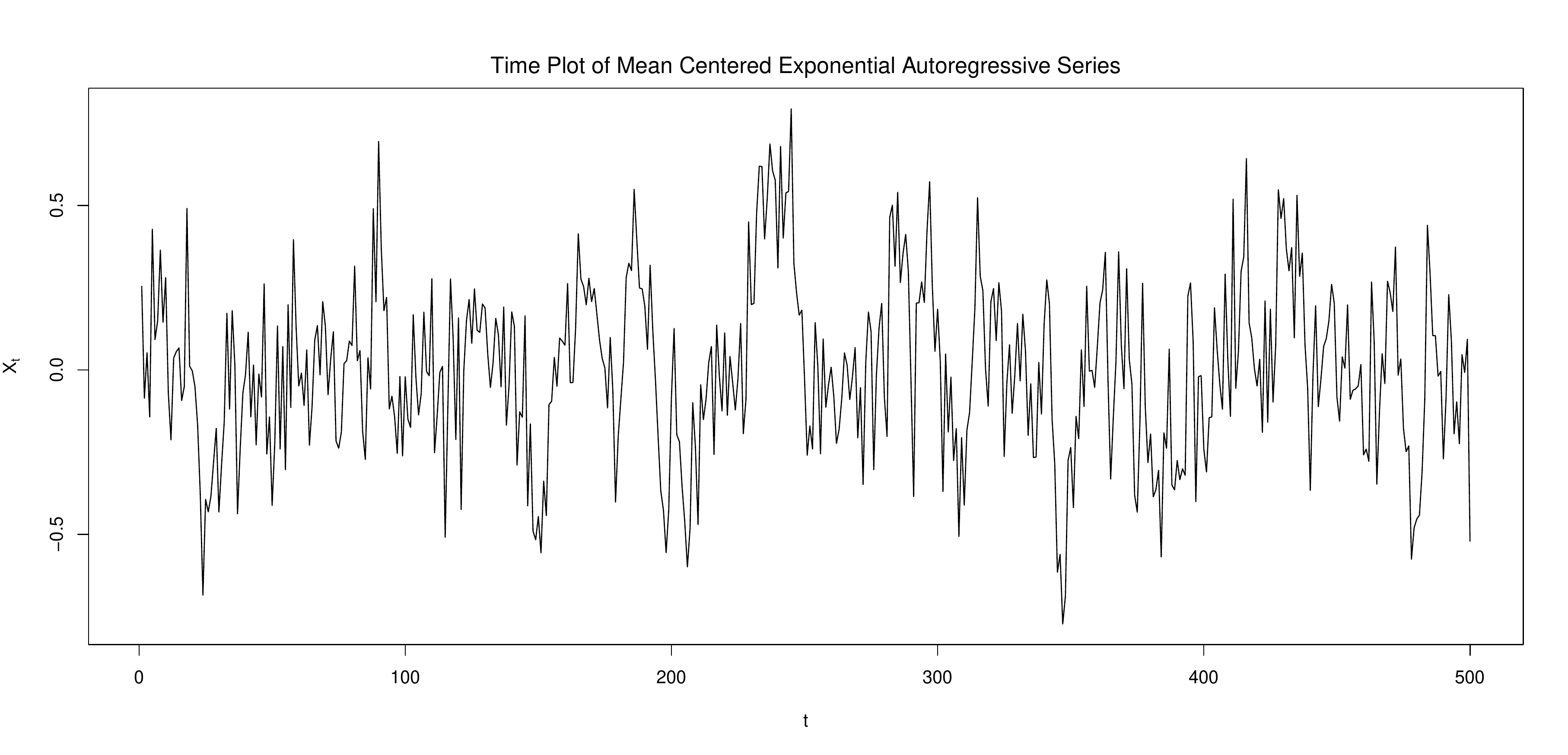}
\caption{Mean-centered realization of length $T = 500$ from an
  EXPAR(2) model given in equation~(\ref{eq:expar}).}
\label{fig:expar}
\end{center}
\end{figure}

To illustrate, consider a series of $T = 500$ observations from the
exponential autoregressive model of order $p = 2$ (EXPAR(2)) given by
\begin{equation}
\label{eq:expar}
X_t = \left\{0.5 - 1.1 e^{-50X_{t-1}^2}\right\}X_{t-1} + 
      \left\{0.3 - 0.5 e^{-50X_{t-1}^2}\right\}X_{t-2} + 0.2\omega_t,
\end{equation}
where the $\omega_t$ are standard normal errors.  A time plot of a
mean-centered realization of length 500 of such a series is given in
Figure~\ref{fig:expar}.
\begin{center}
{\large{Figure~\ref{fig:expar} about here.}}
\end{center}
Since $X_{t-1}$ is the functional variable, and is one of the
autoregressive lags, the model in (\ref{eq:expar}) must be rewritten
and treated as
$$
X_t = m_0\left(X_{t-1}\right) + m_1\left(X_{t-1}\right)X_{t-2} + 0.2\omega_t.
$$
Consequently the functional coefficients of the autoregressive terms
are
$$
m_0(u_t) = 0.5u_t - 1.1 u_t e^{-50u_t^2} \quad {\mbox{and}} \quad
m_1(u_t) = 0.3 - 0.5 e^{-50u_t^2},
$$ 
where $u_t = X_{t-1}$.

To estimate the functional coefficients, begin by accounting for the
variability in the response due to the term $m_0(u_t)$.  Remove
that variability, and use the pseudo-responses to estimate $m_1(u_t)$.
Noting that the maximum lag is 2, we have
\begin{enumerate}
\item For $j = 0$ in equation~(\ref{eq:spline}), fit a spline to the
  mean-centered data.  The result is an estimate $\tilde m_0(u_t)$ of
  $m_0(u_t)$.  Note that the sum of squares in equation~(\ref{eq:SS}) has
  no second sum, since we are considering only a single value of $j, j = 0$;
  that is, equation~(\ref{eq:SS}) reduces to
  $$
  \sum_{t=3}^T \left\{X_t - \sum_{k=0}^{N+1}\lambda_{k,0} b_k(u_t)\right\}^2
  $$
\item Compute pseudo-responses $\hat W_{t,1},~t = 3, \ldots, T$ using
  $$
  W_{t,1} = X_t - \tilde m_0(u_t), \quad t = 3, 4, \ldots T.
  $$
  These pseudo-responses are a proxy for the original realization, but
  with the effect of the $m_0(u_t)$ removed.
\item Fit a kernel regression to the pseudo-responses to get the
  SBK estimate $\hat m_1(u_t)$ of $m_1(u_t)$.  
\end{enumerate}
Repeat the procedure, reversing the roles of $m_0$ and $m_1$.  To get
the coefficients $\lambda_{k,2}, k = 0, 1, \ldots, N+1$ for the spline,
minimize the sums of squares
$$
  \sum_{t=3}^T \left[X_t - \left\{\sum_{k=0}^{N+1}\lambda_{k,0}
     b_k(u_t)\right\}X_{t-2}\right]^2.
$$
The pseudo-reponses $W_{t,2},~t = 3, 4, \ldots, T$ are computed via
$$
  W_{t,2} = X_t - \tilde m_1(u_2)X_{t-2}.
$$

\begin{figure}
\begin{center}
\includegraphics[width=0.9\textwidth]{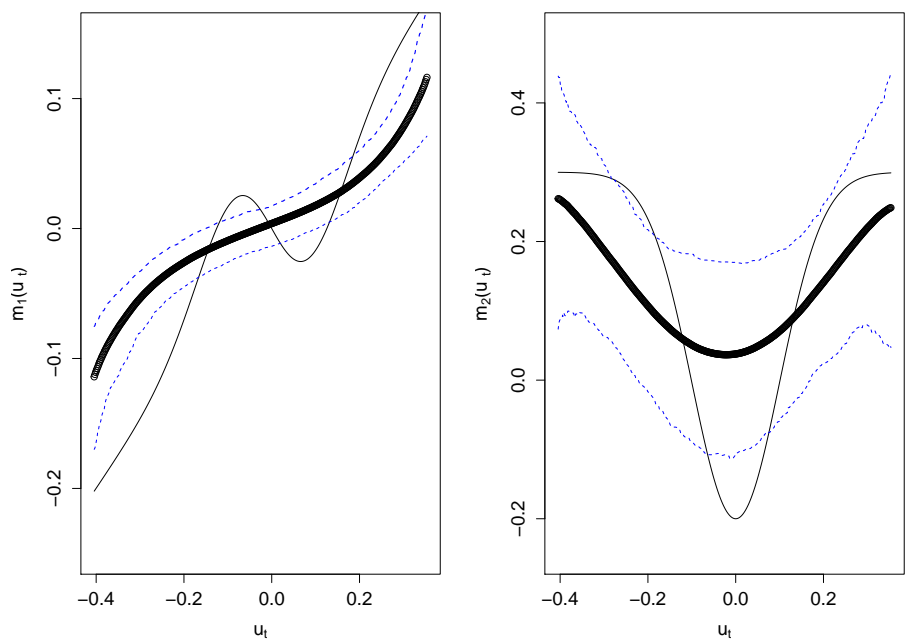}
\caption{Spline-backfitted kernel estimates of the coefficients
  $m_0(u) = 0.5u - 1.1u\exp\{-50u^2\}$ (left panel) and $m_1(u) = 0.3
  - 0.5 \exp\{-50u^2\}$ (right panel).  Heavy curves are the
  estimates $\hat m_0(u)$ and $\hat m_1(u)$; thinner lines are the true
  functions $m_0(u)$ and $m_1(u)$; dashed lines are 95\% confidence
  bands.}
\label{fig:estcoeffs}
\end{center}
\end{figure}

Figure~\ref{fig:estcoeffs} shows the estimation results of a simulated
series from the exponential autoregressive model in
equation~(\ref{eq:expar}) with IID standard normal $\omega_t$ and 500
samples.  The dark curves of dots are the estimated functions, and the
solid (thin) lines are the true functions.  The dashed lines are the
95\% pointwise confidence bands.
\begin{center}
{\large{Figure~\ref{fig:estcoeffs} about here.}}
\end{center}

\subsection{Spatial Modeling for Lattice Data}
\label{sec:spacetime}
For a fixed time $t$, consider a lattice process $Y_s,~s = 1, 2,
\ldots S$.  In this section, to ease notation, the time index $t$ is
suppressed.  Let ${\mathcal{N}}_s$ represent a neighborhood around
location $s$.  The simultaneous autoregressive (SAR) model is defined
as
$$
Y_s = \sum_{{j^\prime}\in\mathcal{N}_s} \beta_{s,{j^\prime}} Y_{j^\prime} + \delta_s,
$$ 
where $\beta_{s,{j^\prime}}$ is a set of coefficients that induces the
spatial autocorrelation between locations ${j^\prime}$ and $s$ in
$\mathcal{N}_s$, and $\delta_s$ are independent, zero-mean, constant
variance errors.  The SAR model was first introduced by
\cite{whittle1954}.  The adjective ``simultaneous'' describes the $S$
autoregressions that occur simultaneously at each data location in the
formulation.  To fit this model in Section~\ref{sec:application}, we
will employ a two nearest neighbor structure to define
$\mathcal{N}_s$.  The model is fitted using maximum likelihood
estimators which are obtained using the {\tt R} package {\tt spdep}
\citep{bivand2013}.

\subsection{Spatio-temporal Modeling}
\label{sec:FCSAR}
We now introduce two spatio-temporal models both of the form in
\eqref{eq:st1}.  The first model we consider uses the noise process
defined in \eqref{eq:st2} and assumes a separable covariance structure
for this process.  The time series structure, $X_{s,t}$, is modeled as
an FCAR model using the SBK method.  Values of $p$ and $d$ in
\eqref{eq:fcar} are allowed to vary between locations, and the spatial
structure, $Y_{s,t}$, is modeled separately using a SAR model at each
time.  By modeling the time and space components separately, we are
implicitly assuming separability.  If this assumption is appropriate,
then the order in which the two models are fit (time-then-space, or
space-then-time) should not matter.

The second spatio-temporal model does not assume separability.
Combining the FCAR($p$) model with a generalized version of the SAR
model, we define the space-time functional coefficient simultaneous
autoregressive (FCSAR) model as
\begin{equation}
\label{eq:fcsar}
Z_{s,t} = \sum_{w=1}^b \sum_{\ell\in\mathcal{N}_s} \beta_{s,\ell,w}Z_{\ell,t-w} + 
  \sum_{k=1}^{p_s} m_{k,s}\left(Z_{s,t-d_s}\right)Z_{s,t-k} + \varepsilon_{s,t},
\end{equation}
where $b$ is the spatial time order for the spatial component in the
model, $\beta_{s,\ell,w}$ is the spatial autocorrelation between
locations $s$ and $\ell$ at a time lag of $w$, $Z_{s,t-d_s}$ is a delay
variable, and $\varepsilon_{s,t}$ are IID with mean zero and constant
variance $\sigma^2_\varepsilon$.  We allow the values of $p_s$ and
$d_s$ to vary among locations.  The FCSAR model is based on the
space-time simultaneously specified autoregressive model of
\cite{woolrich2004}.
\section{Application and discussion}
\label{sec:application}
To illustrate the utility and compare the performance of the proposed
models, we model GHI data at the 1.2 MV La Ola PV plant on the island of Lanai,
Hawaii.  We chose to model GHI rather than POA irradiance to illustrate a more general application of our method.  The La Ola PV plant contains a grid of 12 single-axis tracked arrays arranged in three columns and four rows covering a total area of approximately 250m by 250m. At the time this work was
undertaken, the La Ola data comprised the only available irradiance
data set with concurrent measurements from a regular grid of sensors
across the footprint of a single PV power plant. However, the La Ola data are POA irradiance rather than GHI. Sandia National Laboratories and SunPower Corporation designed an irradiance measurement system in part to study the effects of the movement of cloud shadows across the PV arrays on the power output of the plant
\citep{kuszamaul2010}.  Plane-of-array (POA) irradiance (in
$\text{W}/\text{m}^2$) is measured at the midpoint of each tracking
array using LiCor-200 pyranometers.

Before fitting the models it is necessary to remove the diurnal
trend, a step which we found somewhat difficult.  Clear sky models are
available for removing trends from measured GHI data; a review of some
of these models can be found in \cite{reno2012}.  We set out to use clear sky models to remove the diurnal trend, which would present no great difficulty for measured GHI. We know of no equivalent ``clear-sky'' model for POA irradiance (although, if the
tracking algorithm is known with sufficient precision, such a model could be
assembled by applying a GHI-to-POA translation model, e.g., the DISC
model of \cite{maxwell} to the output of a clear-sky model).  We translated POA irradiance to GHI by assuming the isotropic sky model for the sky diffuse irradiance and using concurrent measurements of diffuse horizontal irradiance (DHI) and direct normal irradiance (DNI) from a nearby rotating shadowband radiometer (RSR) operated
by the National Renewable Energy Laboratory. Because tracker rotations are not measured we estimated the angle of incidence on the modules using a generic algorithm for single-axis tracking \citep{lorenzo2011}.
Even with the use of
measured DHI and DNI, the estimated GHI profiles were not well-matched with
the output of available clear-sky models, and the clear sky models
performed poorly in removing the trend.  Consequently, we removed the
diurnal trend in the estimated GHI by using a local polynomial kernel
regression implemented in the {\tt KernSmooth} package
\citep{wand2012} in the {\tt R} programming software.  

We selected one year (i.e., January 1, 2010 to December 31, 2010) of
POA irradiance measurements, which are recorded every second.  We
observed little to no variability from one irradiance measurement to
the next at one second intervals and consequently reduced the data by
time averaging.  We investigated time-averages of lengths of 30
seconds, 1 minute, 5 minutes, and 10 minutes.  Longer time averages (e.g., 15 and 20 minutes) were also considered but did not appear to be signficantly different from the 10 minute averages. Much of our exploratory work was done using 10-minute averaged data to reduce computational burdens.

\begin{figure}
\begin{center}
\includegraphics[width=\textwidth]{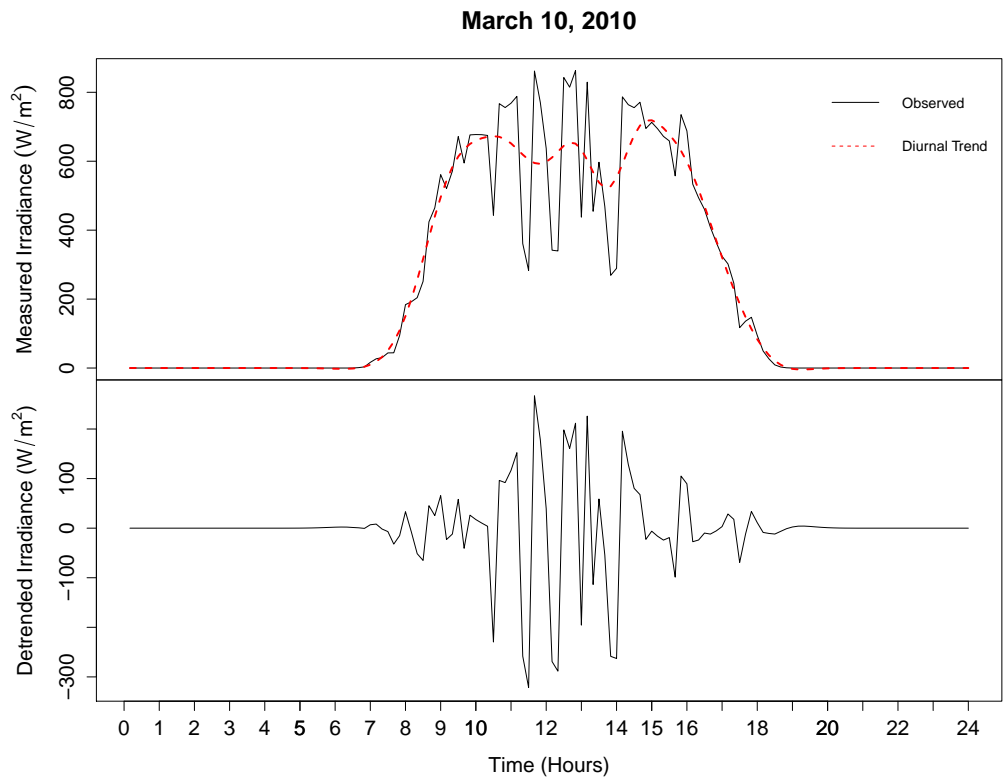}
\caption{Top graph is the time plot of 10-minute averages (solid
  black) of irradiance measurements for March 10 with the local
  polynomial kernel estimate (dashed red) superimposed.  The bottom
  plot is transformed irradiance (residuals after using local
  polynomial kernel regression to remove the diurnal trend).}
\label{fig:remove_trend}
\end{center}
\end{figure}

The top time plot in Figure~\ref{fig:remove_trend} contains the
10-minute time averages of estimated GHI in solid black superimposed
with the local polynomial kernel regression estimate in dashed red for
March 10.  The bottom time plot contains the residuals, hereafter referred to as ``transformed irradiance,'' obtained after removing the diurnal trend by subtracting the kernel fit.
\begin{center}
{\large{Figure~\ref{fig:remove_trend} about here.}}
\end{center}

\begin{figure}
\begin{center}
\subfigure[October 21, clear.]{\label{fig:clear}{%
   \includegraphics[height=0.28\textheight]{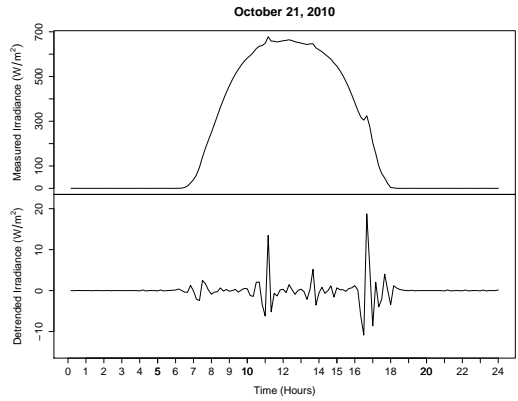}}}
\subfigure[April 1, partly cloudy.]{\label{fig:cloudy}{%
    \includegraphics[height=0.28\textheight]{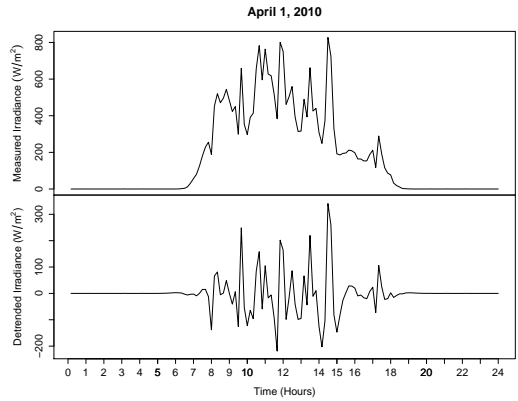}}} \\
\subfigure[August 3, overcast.]{\label{fig:rainy}{%
    \includegraphics[height=0.28\textheight]{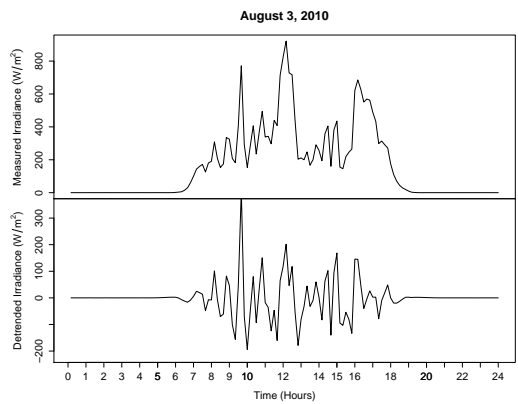}}}
\caption{Time plots of 10-minute time averaged irradiance and of
  transformed irradiance for (a) a clear day, October 21, (b) a partly cloudy day, April 1, and (c) a overcast day, August 3.}
\label{fig:days}
\end{center}
\end{figure}

Having removed the diurnal trend, we next examined a large number of
time plots of GHI to find days with different variability characteristics.  For each day, the weather condition was
classified visually as being in one of three categories: clear, partly
cloudy, and overcast, by the variability and magnitude of GHI.  Figure \ref{fig:days} shows the 10-minute time
averaged irradiance and the transformed irradiance a clear day
(October 21), a partly cloudy day (April 1), and an overcast day
(August 3).
\begin{center}
{\large{Figures~\ref{fig:clear} through~\ref{fig:rainy} about here.}}
\end{center}
For 2010, in Lanai, HI, only six days could be classified as ``clear''
throughout the entire day.  For partly cloudy and overcast conditions,
we found many days.  For both of these weather conditions, six days in 2010 
were randomly selected.

For each selected day, we explored whether assuming space-time covariance separability in \eqref{eq:st1} would be justified.  Using
the separable models we fit the data in two ways: space-then-time and
time-then-space.  If the separability assumption is appropriate, then
the two models are equivalent and should yield similar results.  For
the space-then-time approach, we first fit the SAR model to the 16
sensors for each time, $t$.  We obtain the residuals from the fitted
SAR model, and then apply the FCAR model to each sensor separately.
For the time-then-space model, we first fit the FCAR model to the
detrended irradiance for each of the 16 sensors, and then the
residuals from the fitted FCAR model are fit with the SAR model at
each time point.  For each approach the root mean square errors (RMSEs) (over all sensor locations and times) for eighteen days with three different weather conditions
are found in the first two columns of Table~\ref{tab:rmse}.  For all
days considered in this study, the RMSE for the model that fit space
first is considerably smaller than than when time was fit first.  This
is a strong indication that the assumption of separable covariance
structures is not supported and that nonseparable models should be employed.

For a fixed time $t$, because PV cells are at fixed locations, the
spatial structure can properly be considered a lattice.  Consequently,
for the nonseparable model we fit the FCSAR model in \eqref{eq:fcsar}
for spatial time orders $b = 1, 2$.  The last two columns in Table~\ref{tab:rmse} contain
values of RMSE for these two fits.  The FCSAR model with $b = 2$ has
the smallest RMSE for all 18 days, indicating the best fit among the
models considered. For cloudy and partly cloudy conditions RMSE decreases substantially from $b=1$ to $b=2$ indicating that  a lagged model is needed for greater prediction accuracy.
\begin{center}
{\large{Table~\ref{tab:rmse} about here.}}
\end{center}

\begin{figure}
\begin{center}
\subfigure[October 21, Time-Space.]{\label{fig:oct21ts}{%
  \includegraphics[height=0.25\textheight]{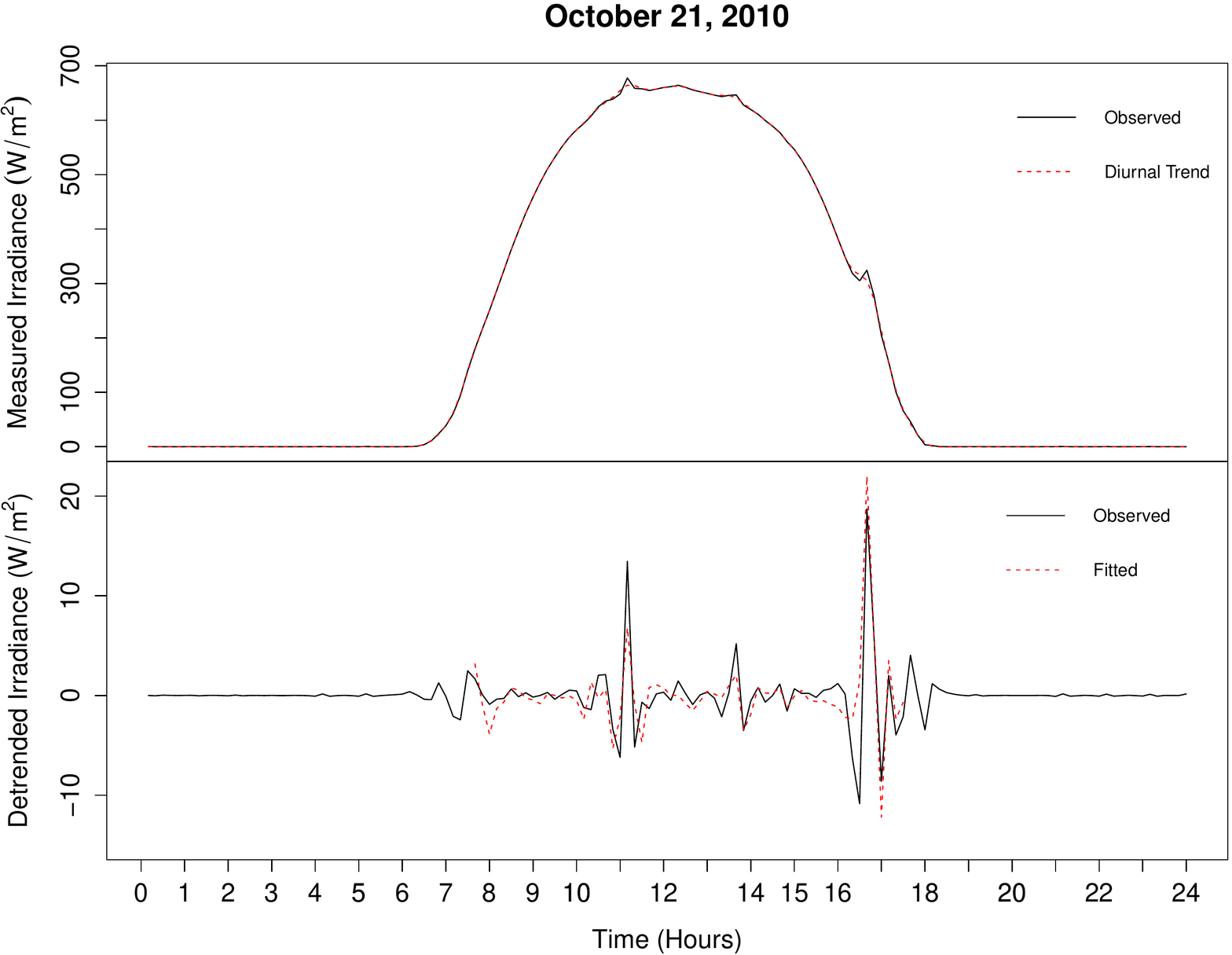}}}
\subfigure[October 21, Space-Time.]{\label{fig:oct21st}{%
  \includegraphics[height=0.25\textheight]{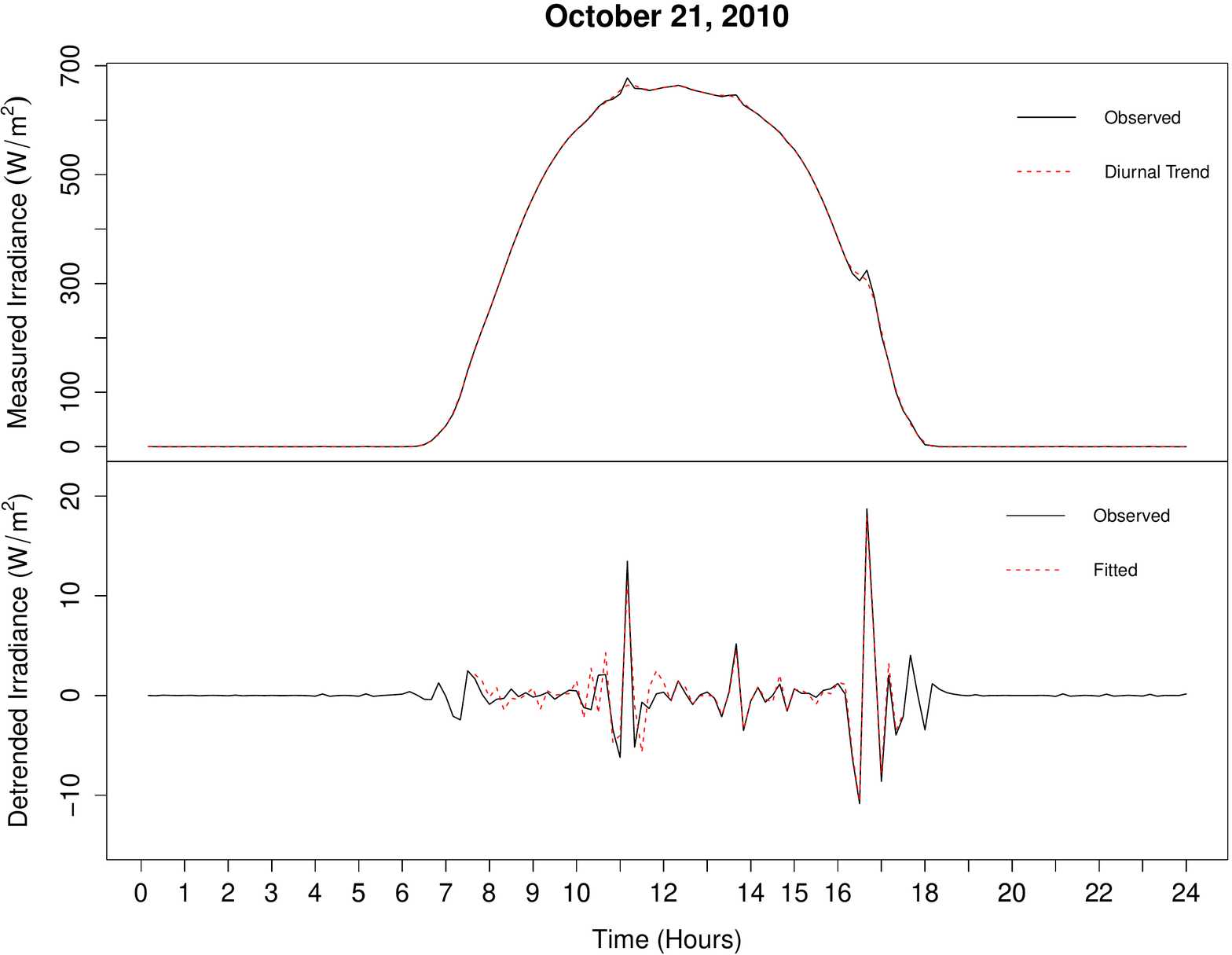}}}
\subfigure[October 21, Nonseparable.]{\label{fig:oct21ns}{%
 \includegraphics[height=0.25\textheight]{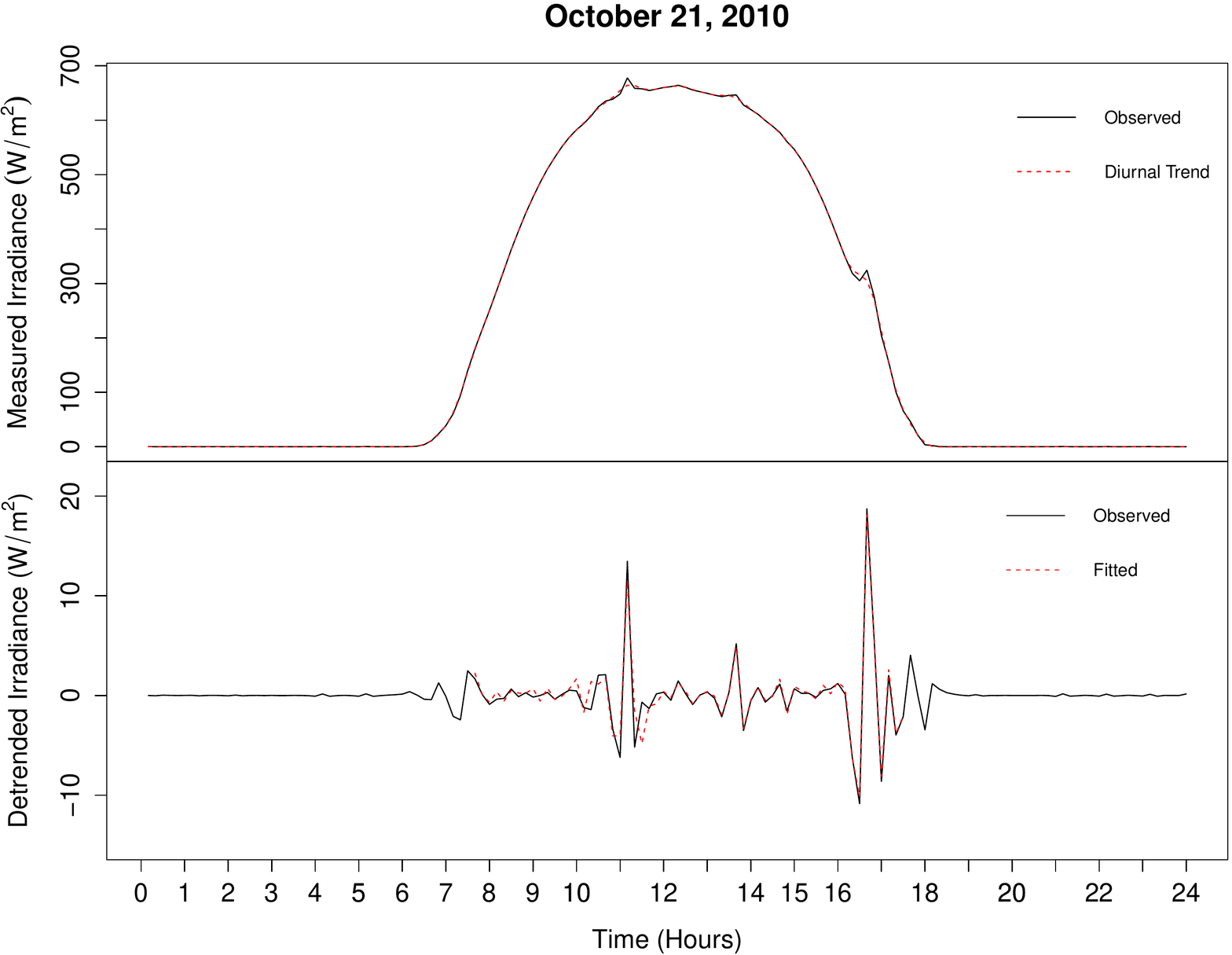}}}
\caption{Time plots of 10-minute time averaged irradiance and of
  transformed irradiance with predicted values superimposed in red for
  a clear day, October 21.  Forecasting was conducted using the (a)
  time-then-space fitting (b) space-then-time fitting (c) nonseparable
  model.}
\label{fig:oct21}
\end{center}
\end{figure}

\begin{figure}
\begin{center}
\subfigure[April 1, Time-Space.]{\label{fig:apr1ts}{%
  \includegraphics[height=0.25\textheight]{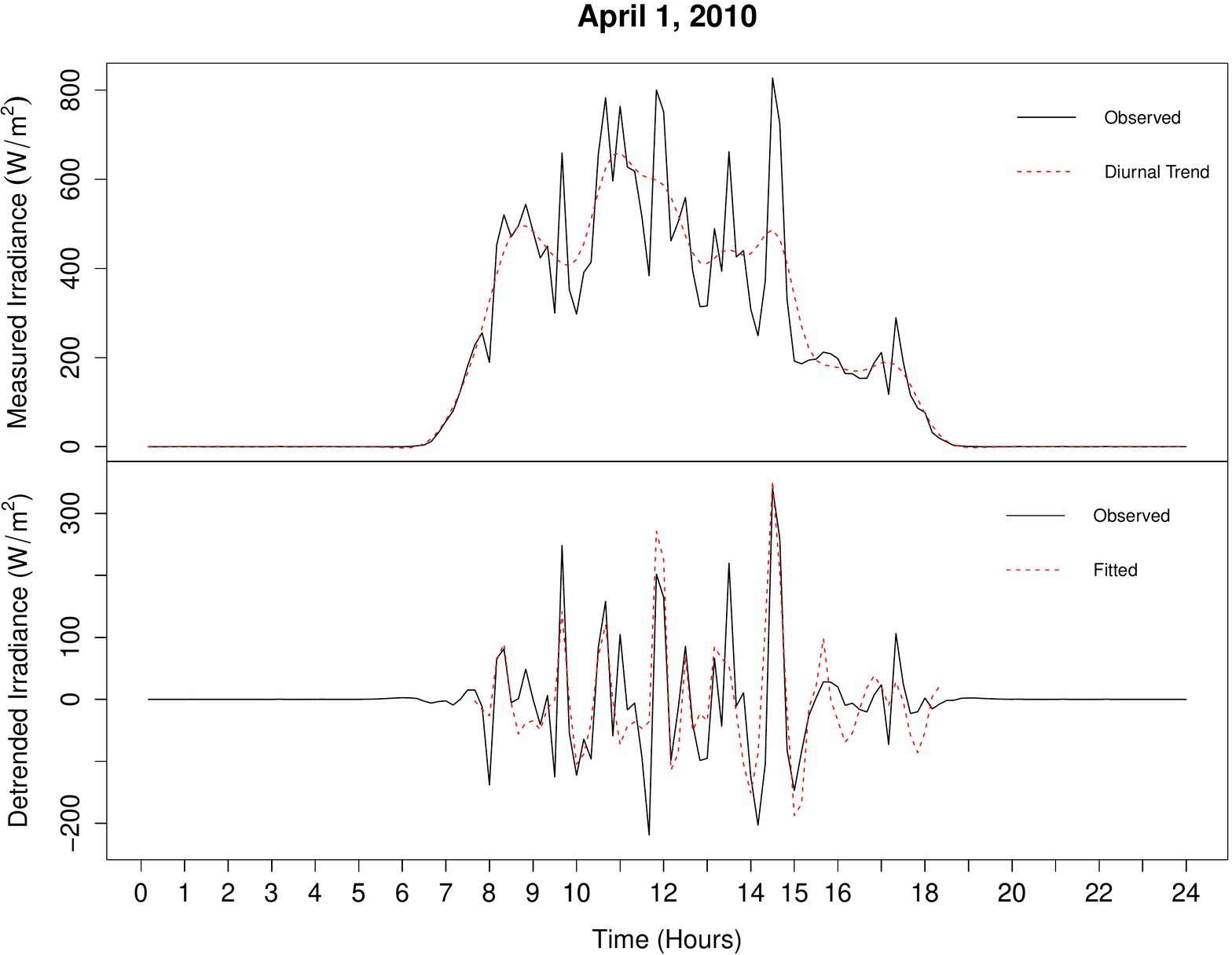}}}
\subfigure[April 1, Space-Time.]{\label{fig:apr1st}{%
  \includegraphics[height=0.25\textheight]{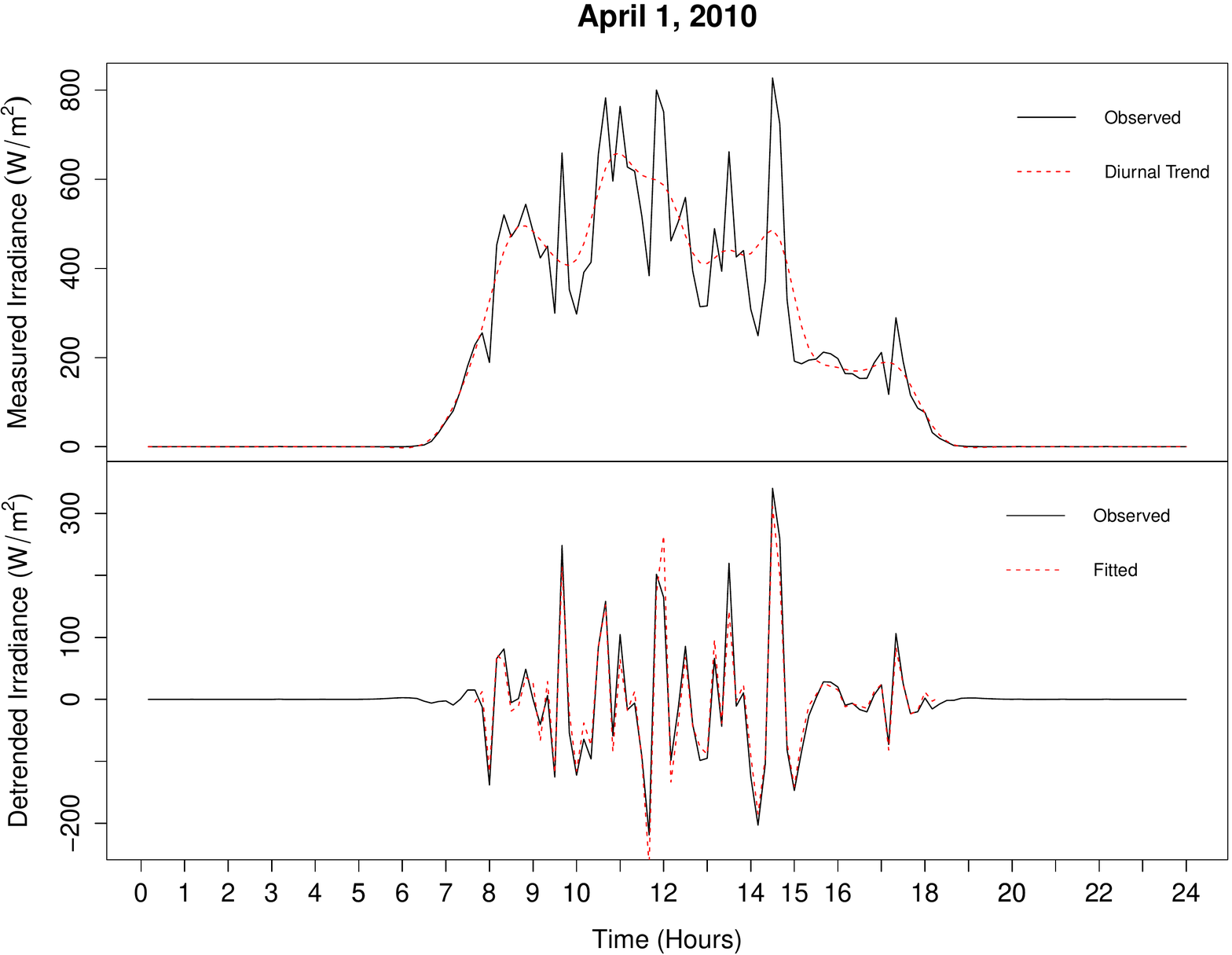}}}
\subfigure[April 1, Nonseparable.]{\label{fig:apr1ns}{%
  \includegraphics[height=0.25\textheight]{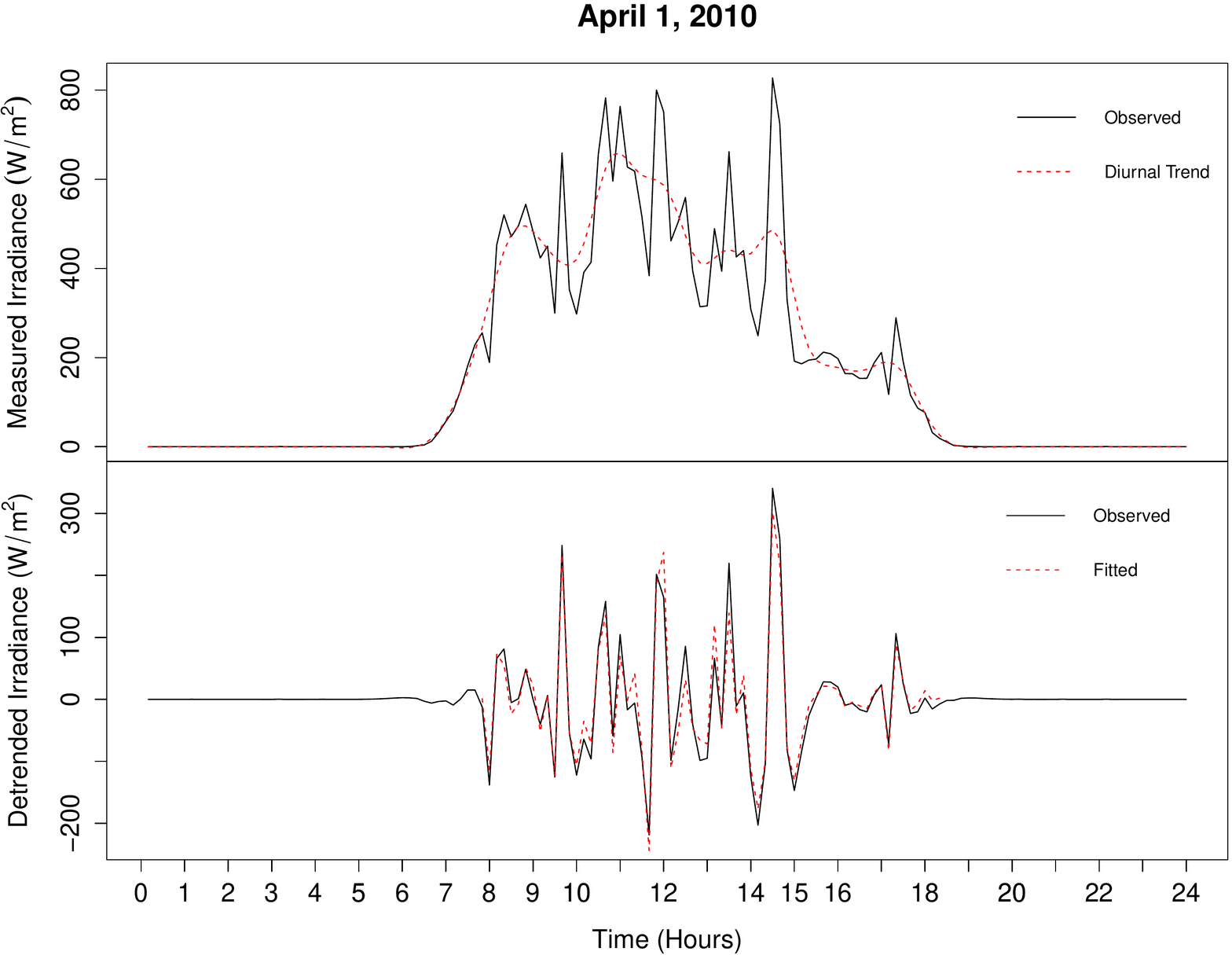}}}
\caption{Time plots of 10-minute time averaged irradiance and of
  transformed irradiance with predicted values superimposed in red for
  a partly cloudy day, April 1.  Forecasting was conducted using the (a)
  time-then-space fitting (b) space-then-time fitting (c) nonseparable model.}
\label{fig:apr1}
\end{center}
\end{figure}

\begin{figure}
\begin{center}
\subfigure[August 3, Time-Space.]{\label{fig:aug3ts}{%
  \includegraphics[height=0.25\textheight]{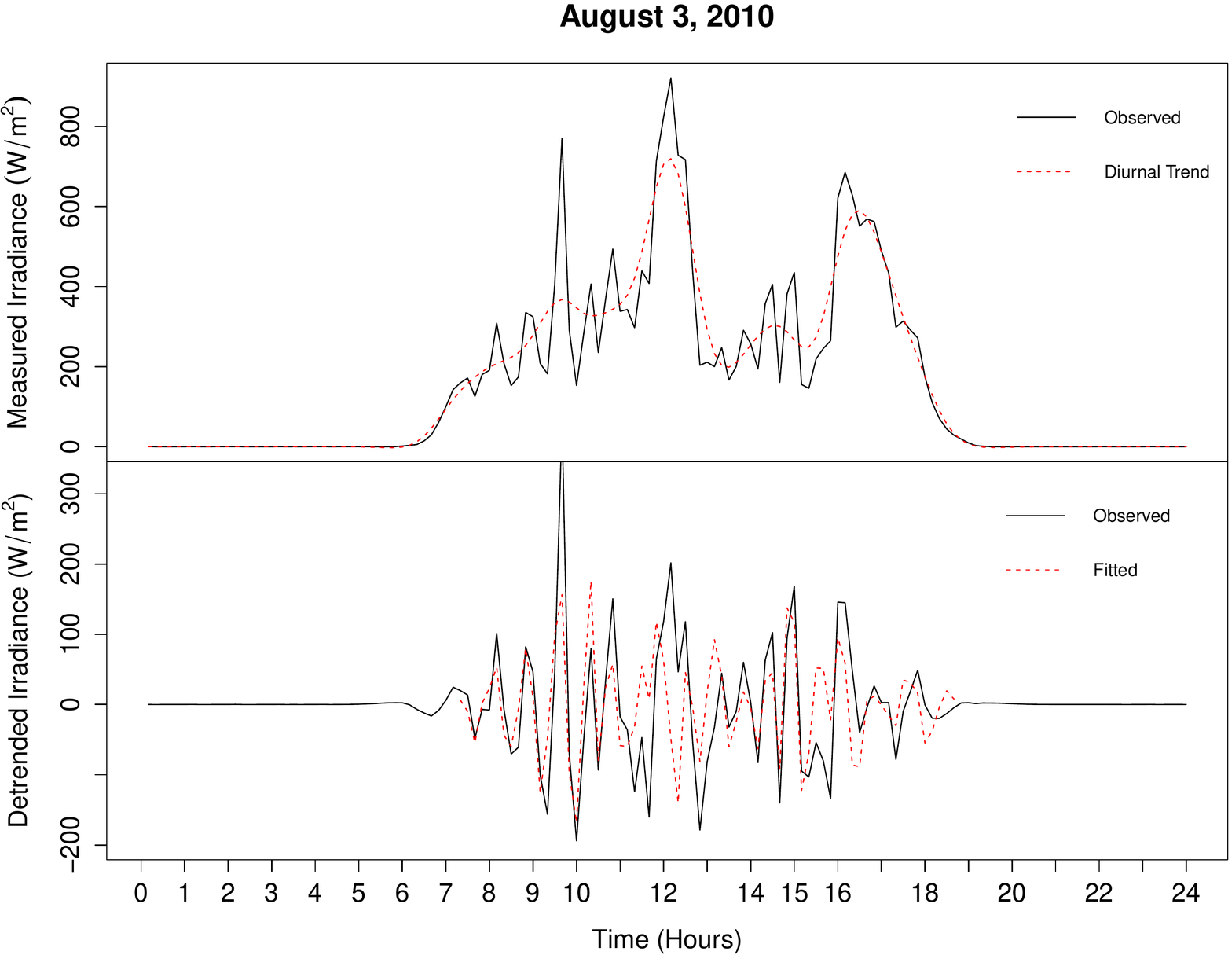}}}
\subfigure[August 3, Space-Time.]{\label{fig:aug3st}{%
  \includegraphics[height=0.25\textheight]{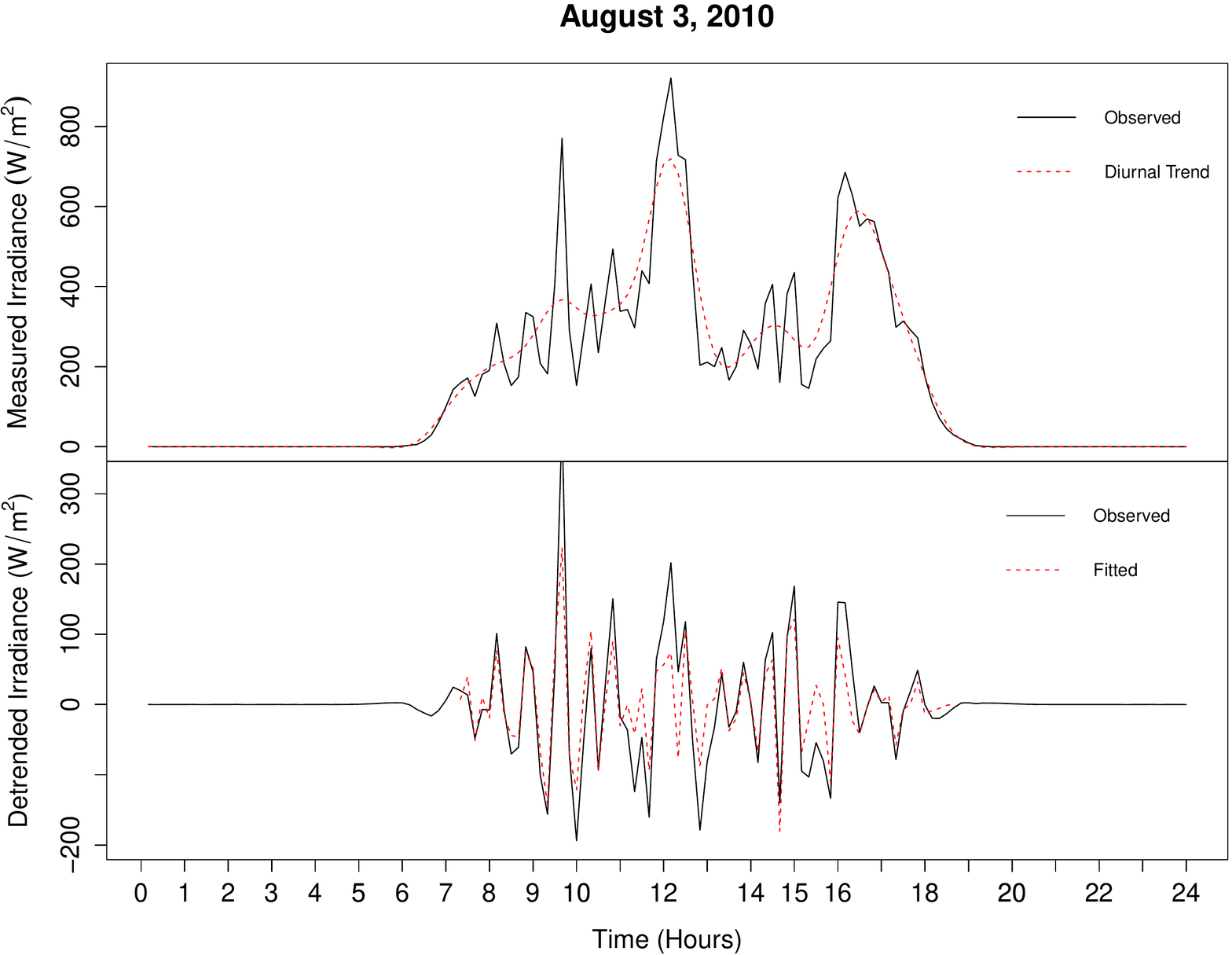}}}
\subfigure[August 3, Nonseparable.]{\label{fig:aug3ns}{%
  \includegraphics[height=0.25\textheight]{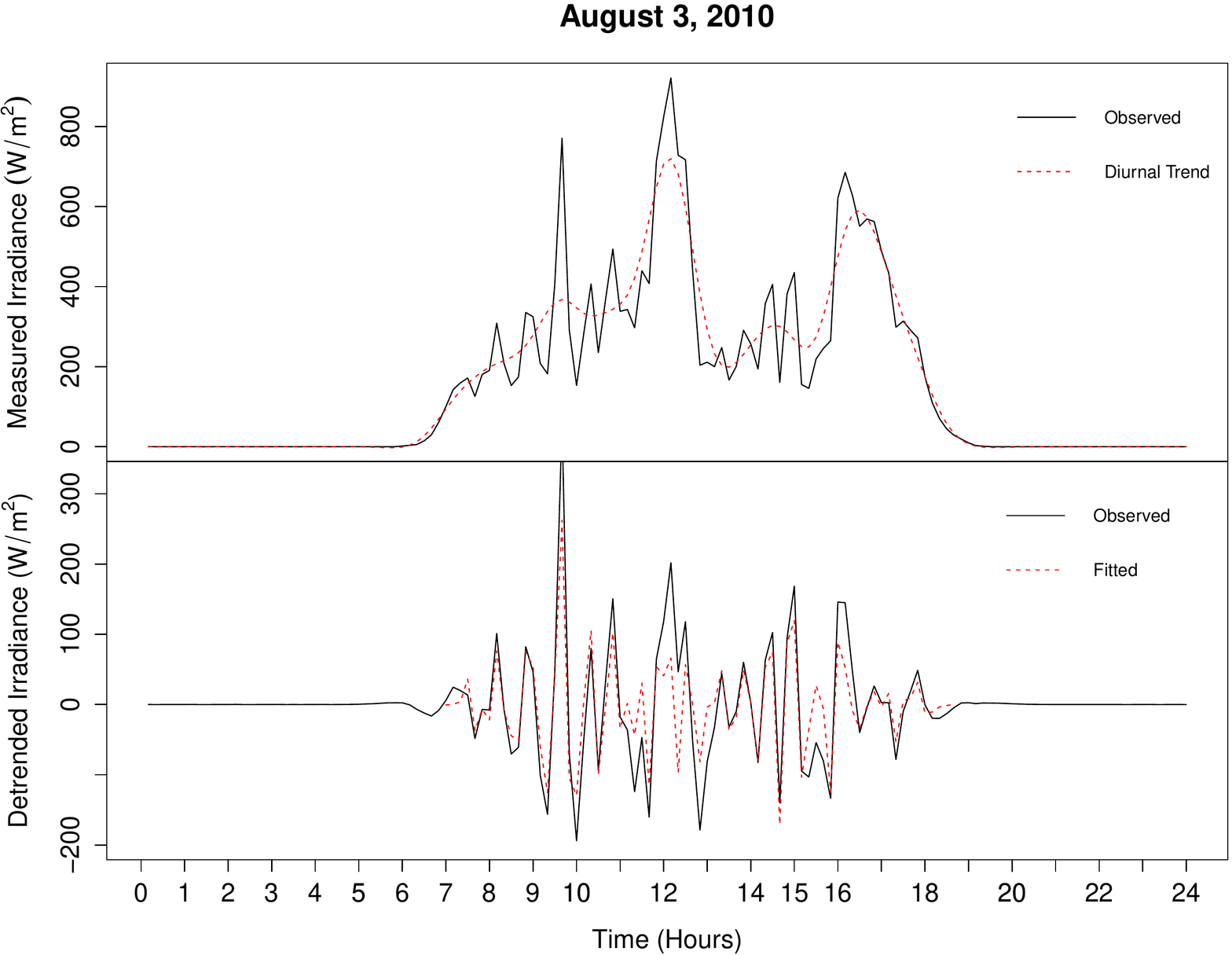}}}
\caption{Time plots of 10-minute time averaged irradiance and of
  transformed irradiance with predicted values superimposed in red for
  an overcast day, August 3.  Forecasting was conducted using the (a)
  time-then-space fitting (b) space-then-time fitting (c) nonseparable
  model.}
\label{fig:aug3}
\end{center}
\end{figure}

Figures~\ref{fig:oct21} through~\ref{fig:aug3} contain six plots, grouped in three pairs.  Each figure displays one sensor location for one day: 
a clear day (October 21, Figure~\ref{fig:oct21});
 a cloudy day (April 1, Figure~\ref{fig:apr1}); and
a partly cloudy day (August 3, Figure~\ref{fig:aug3}).  For any one
pair of plots, the top graph contains the GHI data 
represented by a solid black line, and the modeled GHI represented
by a red dashed line.  The bottom graph contains the detrended GHI data 
(solid black line) and the detrended modeled GHI (red dashed line).
For all three days, the set of two plots labeled (a) were fit using a
separable time-then-space approach; the two sets of plots labeled (b)
were fit using a separable space-then-time approach; and the plots
labeled (c) were fit using the nonseparable FCSAR model with $b = 2$.
\begin{center}
{\large{Figures~\ref{fig:oct21} through~\ref{fig:aug3} about here.}}
\end{center}
The collection of figures illustrates the nonseparable approach yields
the best fit, regardless of the weather conditions, which is in
agreement with minimum RMSE in Table~\ref{tab:rmse}.  However, where
RMSE is an aggregate measure of goodness-of-fit, the plots illustrate
that at individual time points, the goodness-of-fit is uniformly
better for the nonseparable model.


Forecasting the FCSAR model in time is largely dependent on using the SBK method for forecasting the FCAR term in \eqref{eq:fcsar}. In \cite{patrick2015}, methodology is presented for forecasting a FCAR model using the SBK method. In this paper, we examine the performance of forecasting \eqref{eq:fcsar} in space for unobserved locations by using cross-validation.
We simulated unobserved locations by omitting one or several sensors from our data set, and compared FSCAR model performance with a commonly used interpolation technique to judge the potential improvement offered by the FCSAR model. 

Unobserved data are often estimated by interpolating between nearby sensors; one such technique is natural neighbor interpolation which comprises a weighted average with weights determined by a Voronoi partition \citep{sibson1981}. A Voronoi partition divides the space that contains the sensors into regions. Each sensor will have a corresponding region consisting of all points closer to that sensor than to any other. We constructed a Voronoi partition on the set of training sensors along with the location of the missing sensor.  For cross-validation, we took the weighted average of the training sensors where the weights are determined by the size of the regions. This weighted average is used for the prediction for the missing sensors. 

We fit the FCSAR model to the training set of sensors with $b=2$ and using a two nearest neighbor structure for $\mathcal{N}_s$. For each missing sensor, we determined the two nearest neighbors and predicted the irradiance by using the estimated $\beta$'s for those neighbors.

\begin{figure}
\begin{center}
\subfigure[Clear days.]{\label{fig:cv_clear}{%
   \includegraphics[width=0.45\textwidth]{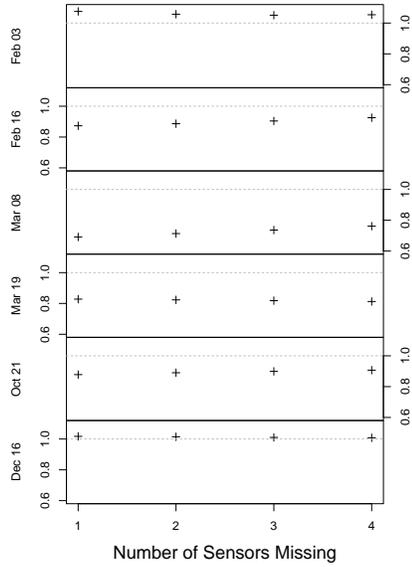}}}
\subfigure[Partly cloudy days.]{\label{fig:cv_cloudy}{%
    \includegraphics[width=0.45\textwidth]{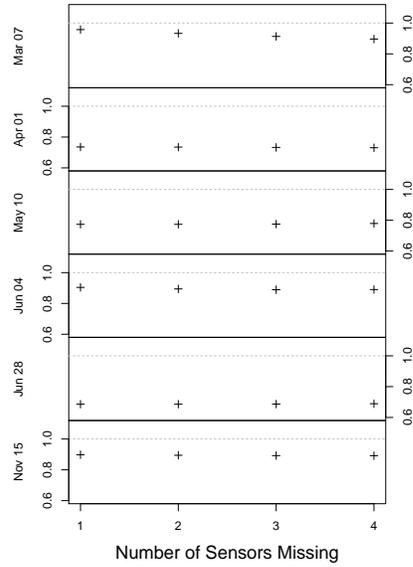}}} \\
\subfigure[Overcast days.]{\label{fig:cv_rainy}{%
    \includegraphics[width=0.45\textwidth]{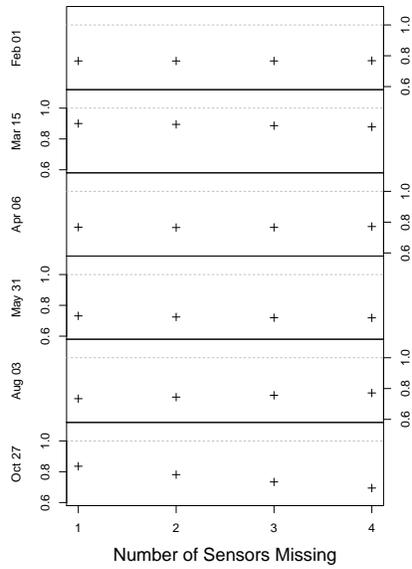}}}
\caption{Plots of the ratios of the RMPE for (a) the clear days, (b) the partly cloudy days, and (c) the overcast days. The ratios are calculated as the RMPE of the FCSAR model divided by the RMPE of linear interpolation.}
\label{fig:cv}
\end{center}
\end{figure}

For our set of 16 sensors, we calculated the predictions with $k=1,2,3,4$ missing sensor locations. For $k>1$, we predicted for each missing location one at a time. We calculated the root mean prediction error (RMPE) as 
$$
RMPE_{\Omega_{i}}=\frac{1}{Tk}\sum_{s\in\Omega_{i}}\left(\sum_{t=1}^{T}\left(\tilde{Z}_{s,t}-Z_{s,t}\right)^{2}\right),
$$
where $\Omega_{i}$ is the $i$th set of $k$ missing sensors, $\tilde{Z}_{s,t}$
is the predicted irradiance for the $s$th sensor at time $t$, and
$Z_{s,t}$ is the observed irradiance. The $RMPE_{\Omega_{i}}$ is calculated for all $K=\left(\begin{array}{c}
16\\
k
\end{array}\right)$ possible combinations of $k$ missing sensors. The mean RMPE is calculated
as
$$
\overline{RMPE}_{k}=\frac{1}{K}\sum_{i=1}^{K}RMPE_{\Omega_{i}}.
$$
To compare the FCSAR model to the interpolation method, we take the ratio
$$
\frac{\overline{RMPE}_{k}\text{ for FCSAR}}{\overline{RMPE}_{k}\text{ for interpolation}}.
$$
The ratios for all 18 days are plotted in Figure \ref{fig:cv}. Ratios less than
one indicates that the FCSAR model performs better at predicting the
missing sensors. All ratios are less than one except for two days
both of which are clear days.

\begin{center}
{\large{Figure~\ref{fig:cv} about here.}}
\end{center}

\begin{figure}
\begin{center}
\subfigure[]{\label{fig:zoom10Oct21}{%
   \includegraphics[width=0.495\textwidth]{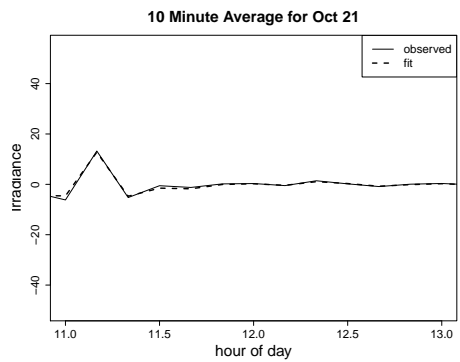}}}
\subfigure[]{\label{fig:zoom5Oct21}{%
   \includegraphics[width=0.495\textwidth]{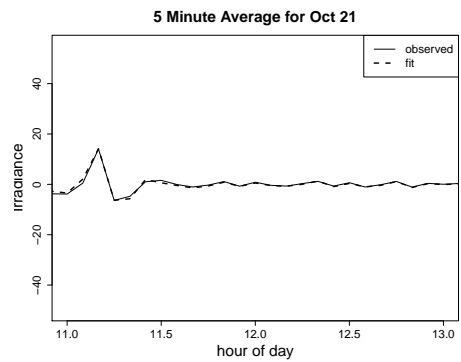}}}\\
\subfigure[]{\label{fig:zoom1Oct21}{%
   \includegraphics[width=0.495\textwidth]{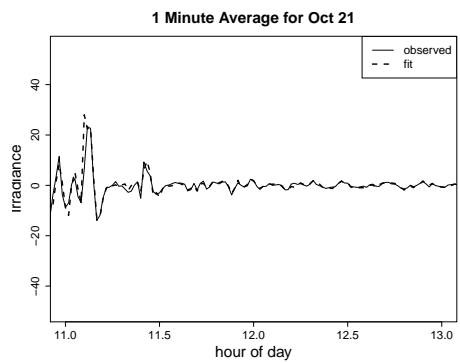}}}
\subfigure[]{\label{fig:zoom30Oct21}{%
   \includegraphics[width=0.495\textwidth]{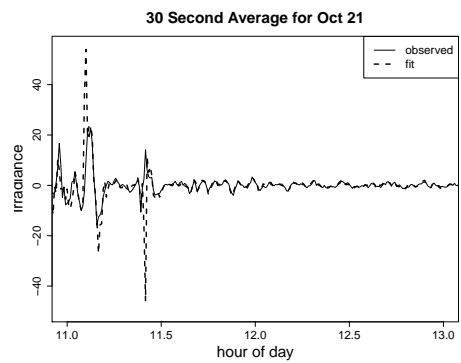}}}
\caption{Plots of the detrended observed irradiance and the fit of the FCSAR model for sensor 1 from 11:00 to 13:00 on October 21 (a clear day). The different plots are for different averages: (a) 10 minutes, (b) 5 minute, (c) 1 minute, and (d) 30 second. }
\label{fig:zoomOct21}
\end{center}
\end{figure}

\begin{figure}
\begin{center}
\subfigure[]{\label{fig:zoom10Apr1}{%
   \includegraphics[width=0.495\textwidth]{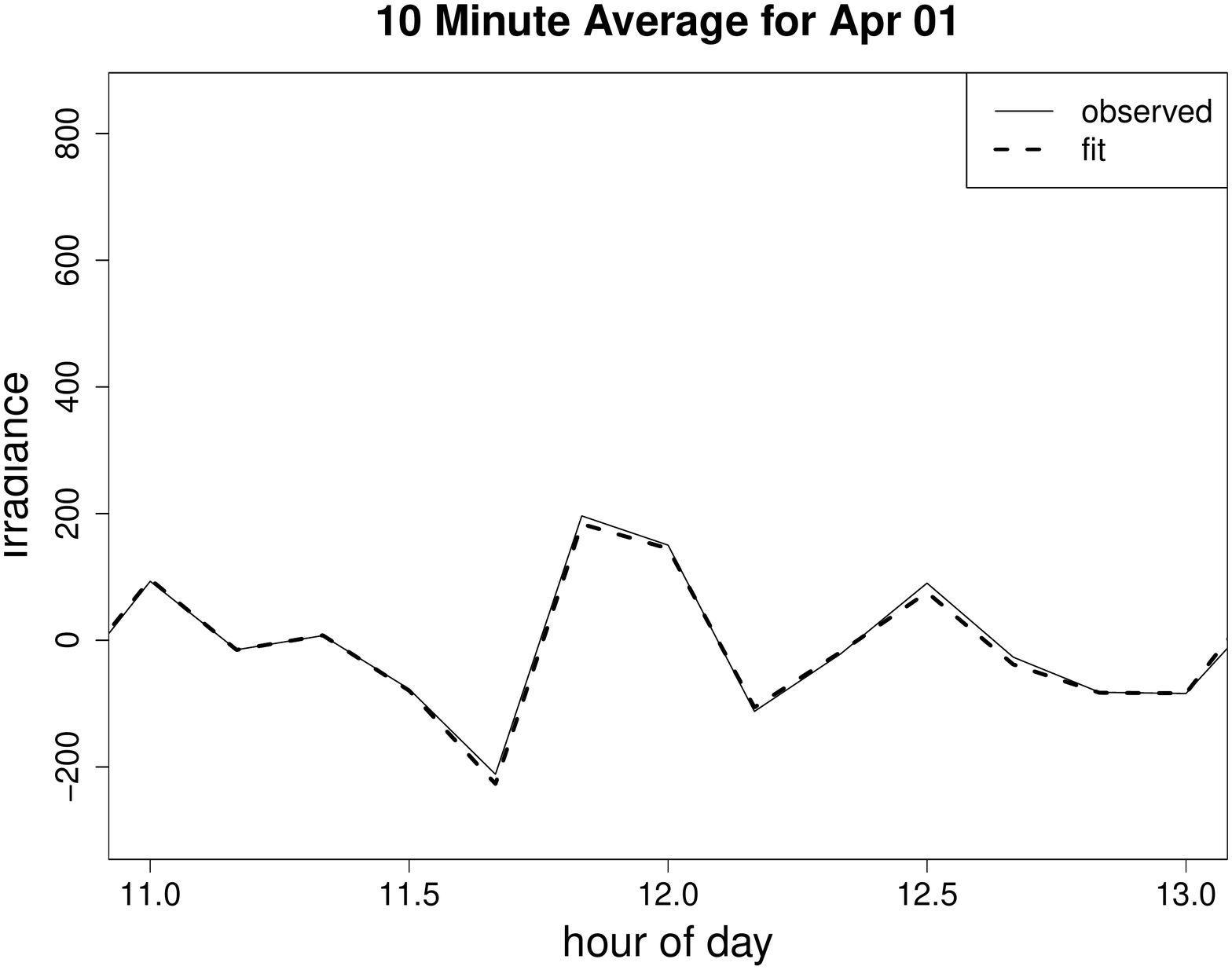}}}
\subfigure[]{\label{fig:zoom5Apr1}{%
   \includegraphics[width=0.495\textwidth]{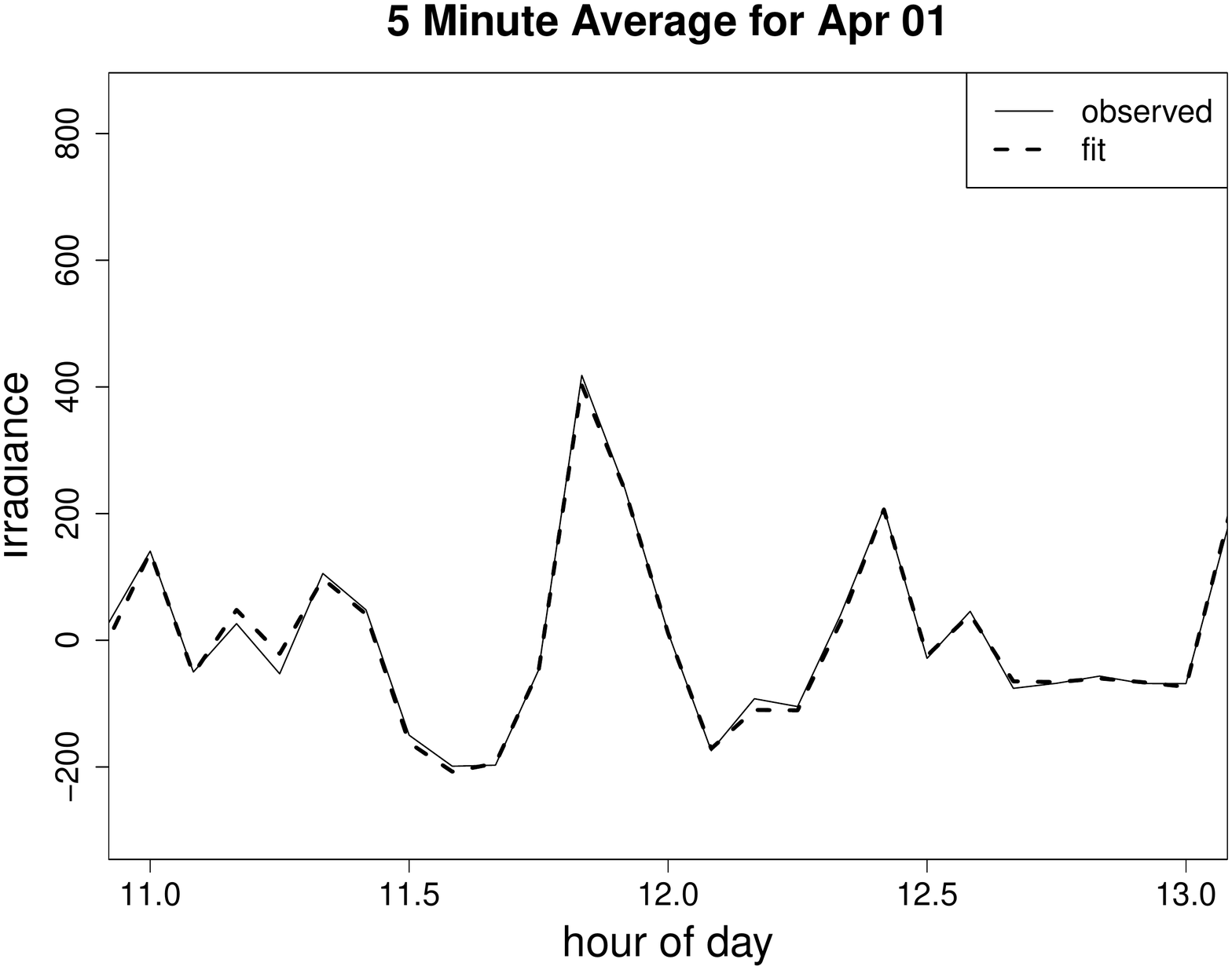}}}\\
\subfigure[]{\label{fig:zoom1Apr1}{%
   \includegraphics[width=0.495\textwidth]{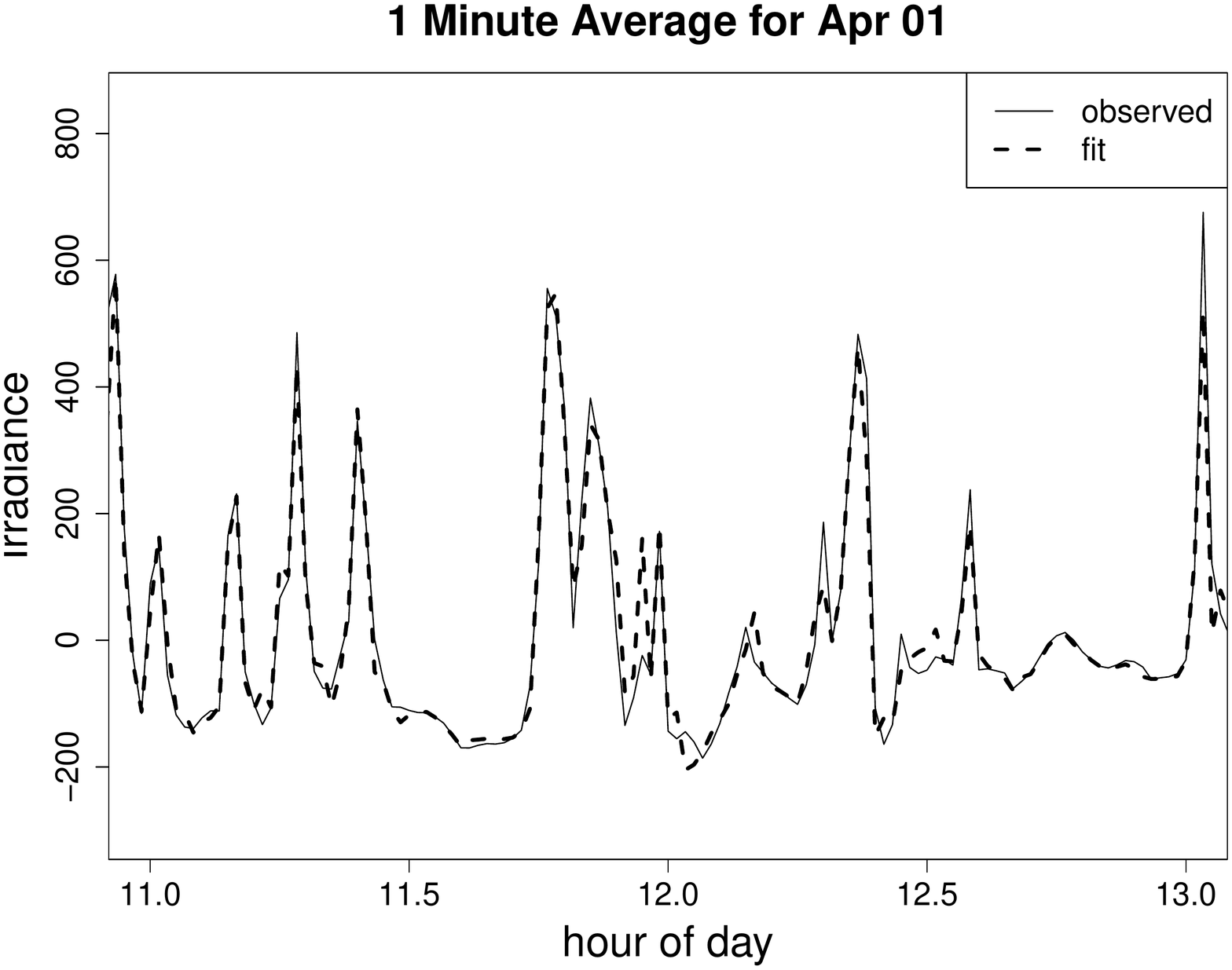}}}
\subfigure[]{\label{fig:zoom30Apr1}{%
   \includegraphics[width=0.495\textwidth]{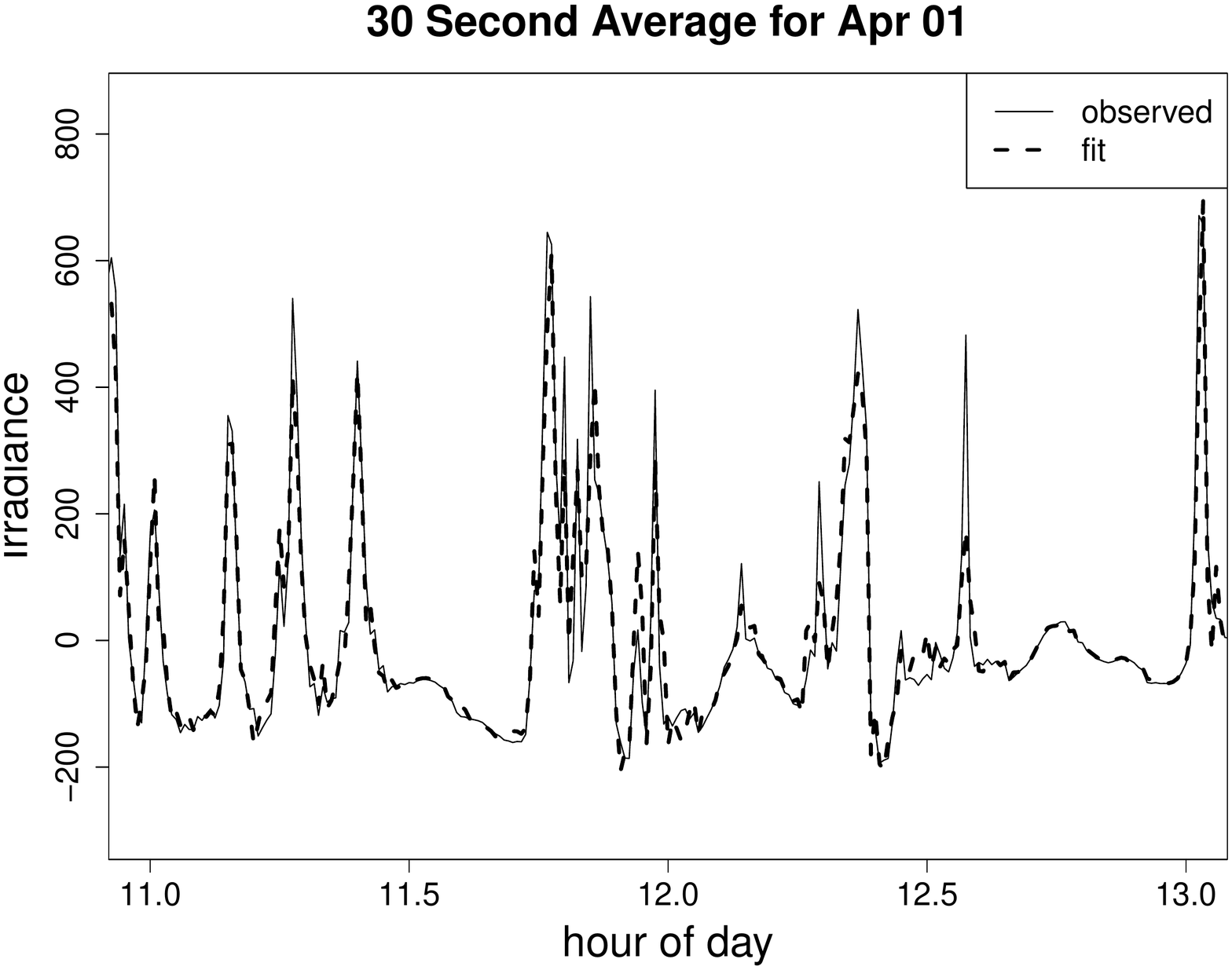}}}
\caption{Plots of the detrended observed irradiance and the fit of the FCSAR model for sensor 1 from 11:00 to 13:00 on April 1 (a partly cloudy day). The different plots are for different averages: (a) 10 minutes, (b) 5 minute, (c) 1 minute, and (d) 30 second. }
\label{fig:zoomApr1}
\end{center}
\end{figure}

\begin{figure}
\begin{center}
\subfigure[]{\label{fig:zoom10Aug3}{%
   \includegraphics[width=0.495\textwidth]{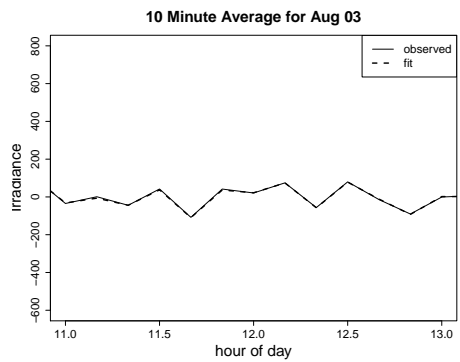}}}
\subfigure[]{\label{fig:zoom5Aug3}{%
   \includegraphics[width=0.495\textwidth]{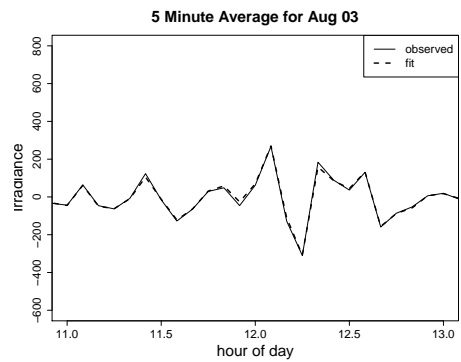}}}\\
\subfigure[]{\label{fig:zoom1Aug3}{%
   \includegraphics[width=0.495\textwidth]{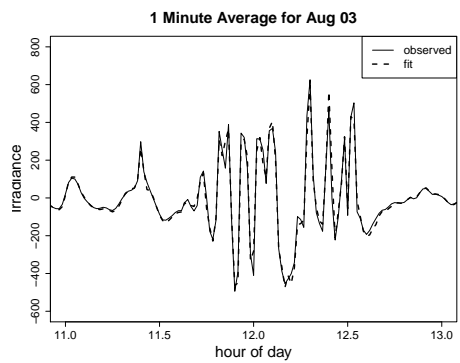}}}
\subfigure[]{\label{fig:zoom30Aug3}{%
   \includegraphics[width=0.495\textwidth]{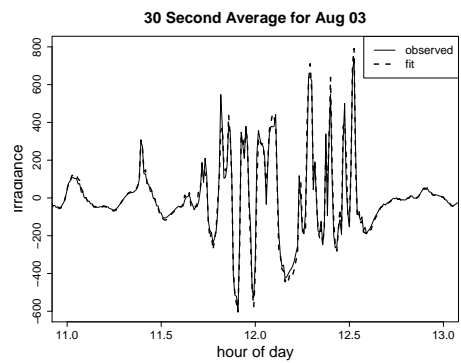}}}
\caption{Plots of the detrended observed irradiance and the fit of the FCSAR model for sensor 1 from 11:00 to 13:00 on August 3 (an overcast day). The different plots are for different averages: (a) 10 minutes, (b) 5 minute, (c) 1 minute, and (d) 30 second. }
\label{fig:zoomAug3}
\end{center}
\end{figure}

To examine the effect of different time averaging windows on the FSCAR model's performance we fit the model for a range of time averaging windows, from 10 minutes down to 30 seconds. For each day, we calculated the model's RMSE as well as the adjusted coefficient of determination $R_{a}^{2}$. The adjusted coefficient of determination $R_{a}^{2}$ quantifies the level of agreement between the data and a fitted model taking into account the number of variables used in the model:
\[
R_{a}^{2}=1-\frac{SS_{\text{Fit}}/\left(TS-\nu_{\text{Fit}}\right)}{SS_{\text{Total}}/\left(TS\right)},
\]
where
\[
SS_{\text{Fit}}=\sum_{s=1}^{S}\sum_{t=1}^{T}\left(\tilde{Z}_{s,t}-Z_{s,t}\right)^{2},
\]
\[
SS_{\text{Total}}=\sum_{s=1}^{S}\sum_{t=1}^{T}\left(Z_{s,t}-\overline{Z}\right)^{2},
\]
\[
\overline{Z}=\frac{1}{TS}\sum_{s=1}^{S}\sum_{t=1}^{T}Z_{s,t},
\]
$\tilde{Z}_{s,t}$ is the predicted irradiance for the $s$th sensor
at time $t$, $Z_{s,t}$ is the observed irradiance, and $\nu_{\text{Fit}}$
is the number of parameters used in the fit. Since we are using kernel
regression to fit the time series, we must estimate the number of
parameters associated with that regression. For the SBK estimate of
the $k$th coefficient function in \eqref{eq:fcsar}, the effective number of parameters
is the trace of the smoother matrix 
\[
\left(\begin{array}{c}
1\\
0
\end{array}\right)\left(\frac{1}{T}\textbf{C}^{\prime}\textbf{M}\textbf{C}\right)^{-1}\frac{1}{T}\textbf{C}^{\prime}\textbf{M}
\]
in \eqref{eq:kernelfit}  \citep[see][]{hastie1990, cai2000}. We calculate the total
number of parameters as the sum of the parameters in the first double
sum in \eqref{eq:fcsar} plus the sum of the effective number of parameters for
the FCAR term.  The values of $R^2_{a}$ and RMSE are shown in Table \ref{tab:comp}. We show the fits of three days for the different time averages in Figures \ref{fig:zoomOct21}-\ref{fig:zoomAug3}.

\begin{center}
{\large{Table~\ref{tab:comp} about here.}}
\end{center}

\begin{figure}
\begin{center}
\subfigure[]{\label{fig:data_boxplot}{%
   \includegraphics[width=0.75\textwidth]{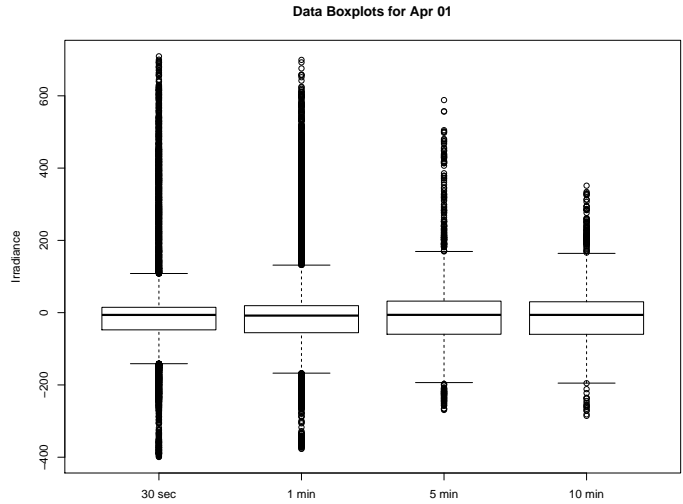}}}\\
\subfigure[]{\label{fig:res_boxplot}{%
   \includegraphics[width=0.75\textwidth]{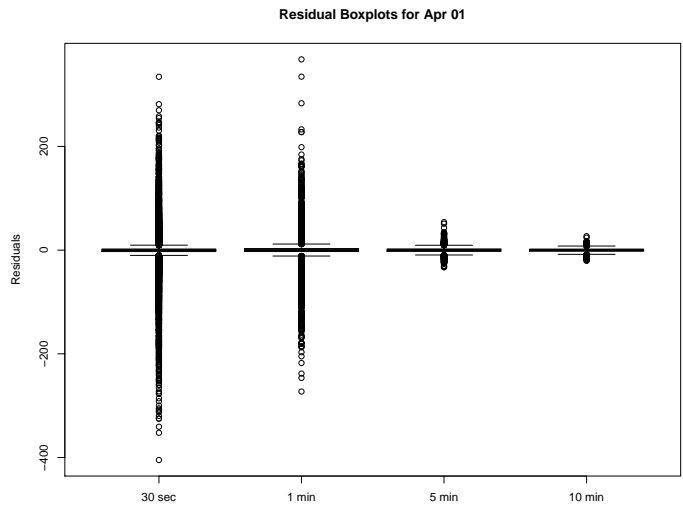}}}
\caption{Boxplots for (a) the observed detrended data and (b) the residuals of the fit of the FCSAR model with $b=2$ for April 1 (a partly cloudy day). }
\label{fig:boxplots}
\end{center}
\end{figure}

Table \ref{tab:comp} shows that as the time averaging window decreases, RMSE increases and and $R_{a}^{2}$ decreases, both indicating increasing disagreement between data and model. However, as the time averaging window decreases, variance in time averaged data at any individual location increases substantially (Figures \ref{fig:zoomOct21}-\ref{fig:zoomAug3}). As ramps in the data increase in both magnitude and frequency the largest residuals of the fitted model also increase. Similar patterns are evident in the spatially-averaged data. Figure~\ref{fig:boxplots} compares distributions for the spatially averaged detrended irradiance data and corresponding distributions residuals for the fitted FSCAR model for a partly cloudy day. As the averaging window decreases., outliers increase in both the data and the model residuals also increase, leading to the increasing RMSE and decreasing $R^2_{a}$ evident in Table \ref{tab:comp}. However, the FCSAR model continues to fit the bulk of the data equally well across all time averaging windows, as is demonstrated by the relatively constant boxes and whiskers across the different time averages. Thus, the FCSAR model follows time averaged data equally well for various averaging windows. 

\begin{center}
{\large{Figures~\ref{fig:zoomOct21} through~\ref{fig:zoomAug3} about here.}}
\end{center}

\begin{center}
{\large{Figure~\ref{fig:boxplots} about here.}}
\end{center}

\section{Conclusion}
\label{sec:conclusion}
We have presented a novel nonseparable spatio-temporal model for GHI data. This approach, termed the FCSAR model, outperforms a natural neighbor interpolation when predicting GHI at unobserved locations over the footprint of a PV system. We compared the nonseparable FCSAR model with simpler, separable models, and find little support for models that assume a separable covariance structure.  The FSCAR model integrates an FCAR form for the time series component of the model and a SAR form for the spatial component. The FCAR($p$) form of the time series component of our nonseparable model makes the FCSAR model flexible and reliable, and may be suitable for fitting irradiance data in general.  Currently, the model is fit separately on each day. Further research will consider validating the fitted models by comparing predicted aggregate irradiance with generated power for a much larger solar power plant than La Ola.  Future work may also explore adding a weather condition covariate that will allow the model to be fit over days with different weather conditions, by permitting the coefficient functions in the time series structure to vary based on weather condition.

\section*{Acknowledgments}
The research was performed under contract (PO 1303122) with Sandia
National Laboratories, a multi-program laboratory managed and operated
by Sandia Corporation, a wholly owned subsidiary of Lockheed Martin
Corporation, for the U.S.~Department of Energy's National Nuclear
Security Administration under contract DE-AC04-94AL85000.
Dr.~Patrick's work was largely completed as a part of his dissertation
work while he was at Baylor University.  The authors thank Justin
Sims for his help in creating the graphs provided in the figures
throughout the manuscript.
\newpage
\bibliographystyle{model2-names}
\bibliography{references}

\newpage
\newpage
\newpage
\newpage
\newpage
\begin{table}
\begin{center}
\caption{Root mean squared error (RMSE) for the four spatio-temporal
  models of the days with clear, partly cloudy, and overcast conditions.
  Columns S-T and T-S contain the RMSE for the separable
  spatio-temporal models.  Column S-T contains RMSE for data with the
  spatial component fit first, then time; Column T-S contains RMSE
  with the time component fit first, then space.  The last two
  columns contain RMSE for the nonseparable FCSAR model with spatial
  time orders $b = 1$ and $b = 2$, respectively.}
\label{tab:rmse}
\begin{tabular}{|c|l|rr|rr|} \hline
  & & \multicolumn{2}{|c|}{Separable} & \multicolumn{2}{c|}{FCSAR} \\ \cline{3-6}
Condition 		& Date    & S-T   & T-S    & $b = 1$ & $b = 2$ \\ \hline
Clear     		& Feb.~3  &  0.36 &   2.45 &  0.36	 &  0.16 \\
          		& Feb.~16 &  3.57 &  14.98 &  2.18   &  1.36 \\
          		& Mar.~18 &  0.34 &   2.64 &  0.34   &  0.22 \\
          		& Mar.~19 &  0.32 &   6.17 &  0.35   &  0.22 \\
         		& Oct.~21 &  0.67 &   4.12 &  0.55   &  0.42 \\
				& Dec.~16 &  0.97 &   6.60 &  0.83   &  0.44 \\ \hline 
Partly Cloudy   & Mar.~7  &  6.88 &  63.74 &  5.49   &  3.35 \\
                & Apr.~1  &  7.61 &  99.56 &  5.47   &  4.51 \\
                & May 10  &  5.91 &  56.86 &  4.90   &  3.72 \\
                & June 4  &  8.46 &  51.58 &  4.84   &  3.36 \\
                & June 28 &  3.63 &  55.10 &  2.39   &  1.76 \\
                & Nov.~15 & 10.39 &  58.24 &  7.14   &  6.00 \\ \hline  
Overcast        & Feb.~1  &  7.52 &  62.61 &  6.20   &  4.88 \\
                & Mar.~15 & 12.82 & 118.23 &  9.12   &  5.50 \\
                & Apr.~6  &  5.66 &  48.15 &  2.97   &  2.43 \\
                & May 31  & 11.44 &  91.35 &  5.92   &  5.08 \\
                & Aug.~3  &  5.63 &  65.81 &  3.93   &  2.82 \\
                & Oct.~27 &  4.07 &  94.30 &  3.10   &  1.89 \\ \hline
\end{tabular}
\end{center}
\end{table}
\newpage
\newpage
\newpage
\newpage
\begin{landscape}
\begin{table}
\begin{center}
\caption{Root mean squared error (RMSE) for the nonseparable spatio-temporal
  models with $b=2$ of the days with clear, partly cloudy, and overcast conditions. The columns are the RMSE for data at 30-second, 1-minute, 5-minute, and 10-minute averages. The values in parenthesis are the $R^2_{a}$. 
  }
\label{tab:comp}
\begin{tabular}{|c|l|rrrr|} \hline
Condition       & Date    & 30-sec   & 1-min  & 5-min   & 10-min       \\ \hline
Clear     		& Feb.~3  &  0.38 (0.990)   &   0.30 (0.993) &  0.19  (0.999) &  0.16 (0.999)       \\
          		& Feb.~16 &  4.98 (0.952)   &   3.68 (0.970) &  1.82  (0.988) &  1.36 (0.963)       \\     
          		& Mar.~18 &  1.79 (0.970)   &   0.99 (0.989) &  0.42  (0.999) &  0.22 (0.992)       \\
          		& Mar.~19 &  0.81 (0.999)   &   0.63 (0.998) &  0.32  (0.952) &  0.22 (0.993)       \\
         		& Oct.~21 &  3.94 (0.846)   &   2.21 (0.926) &  0.67  (0.991) &  0.42 (0.983)       \\
				& Dec.~16 &  0.67 (0.998)   &   0.61 (0.998) &  0.49  (0.952) &  0.44 (0.993)       \\ \hline 
Partly Cloudy   & Mar.~7  & 25.66 (0.920)   &  17.68 (0.960) &  5.42  (0.991) &  3.35 (0.997)       \\
                & Apr.~1  & 31.75 (0.932)   &  24.65 (0.960) &  6.99  (0.996) &  4.51 (0.998)       \\
                & May 10  & 25.91 (0.927)   &  19.68 (0.954) &  7.70  (0.987) &  3.72 (0.999)       \\
                & June 4  & 22.05 (0.907)   &  20.60 (0.920) &  7.21  (0.986) &  3.36 (0.995)       \\
                & June 28 & 13.85 (0.935)   &  10.12 (0.966) &  3.88  (0.994) &  1.76 (0.999)       \\
                & Nov.~15 & 27.69 (0.908)   &  22.56 (0.935) &  9.24  (0.979) &  6.00 (0.987)       \\ \hline
Overcast        & Feb.~1  & 26.46 (0.899)   &  24.12 (0.917) &  9.27  (0.982) &  4.88 (0.993)       \\
                & Mar.~15 & 27.85 (0.938)   &  20.29 (0.968) &  6.05  (0.996) &  5.50 (0.997)       \\
                & Apr.~6  & 24.22 (0.883)   &  19.44 (0.930) &  4.01  (0.997) &  2.43 (0.997)       \\
                & May 31  & 42.26 (0.899)   &  30.26 (0.943) &  9.60  (0.988) &  5.08 (0.997)       \\
                & Aug.~3  & 25.92 (0.948)   &  20.59 (0.966) &  6.15  (0.995) &  2.82 (0.998)       \\
                & Oct.~27 &  7.20 (0.977)   &  5.52  (0.989) &  2.61  (0.998) &  1.89 (0.999)       \\ \hline
\end{tabular}
\end{center}
\end{table}
\end{landscape}
\newpage
\newpage
\newpage
\newpage
\newpage
\end{document}